\documentclass[letterpaper,titlepage,11pt]{article}
\pdfoutput=1
\usepackage[colorlinks=true,urlcolor=blue,citecolor=magenta]{hyperref}
\usepackage{amssymb,amsmath,amsfonts}
\usepackage{epsfig}
\usepackage{epsfig}
\usepackage{graphicx}
\usepackage{epstopdf}
\usepackage{caption}
\usepackage{subcaption}
\def\clock{{\count0=\time
           \divide\count0 60
           \ifnum\count0<10 0\fi\the\count0
           \multiply\count0 -60 \advance\count0 \time
           :\ifnum\count0<10 0\fi \the\count0
         }}
\newcommand{\timestamp}{{\small\vbox{\hbox{\tt\jobname.tex}
\hbox{\the\day/\the\month/\the\year, \clock}}}}

\newcommand{\CS}{\mathcal{S}}

\newcommand{\CW}{\mathcal{W}}
\newcommand{\CV}{\mathcal{V}}

\newcommand{\hT}{\hat{T}}

\newcommand{\spa}{\ , \ \ }

\newcommand{\be}{\begin{eqnarray}}
\newcommand{\ee}{\end{eqnarray}}
\newcommand{\beq}{\begin{eqnarray}}
\newcommand{\eeq}{\end{eqnarray}}

\newcommand{\beqa}{\begin{eqnarray}}
\newcommand{\eeqa}{\end{eqnarray}}

\newcommand{\bk}{\mathbf{k}}

\let\oldsqrt\sqrt
\def\sqrt{\mathpalette\DHLhksqrt}
\def\DHLhksqrt#1#2{%
\setbox0=\hbox{$#1\oldsqrt{#2\,}$}\dimen0=\ht0
\advance\dimen0-0.2\ht0
\setbox2=\hbox{\vrule height\ht0 depth -\dimen0}%
{\box0\lower0.4pt\box2}}

\newcommand{\ads}{\text{AdS}}

\numberwithin{equation}{section}

\begin{document}

\begin{titlepage}
\rightline{\vbox{   \phantom{ghost} }}

 \vskip 1.8 cm

\centerline{\Huge \bf Null-Wave Giant Gravitons from}
\vskip .3cm
\centerline{\Huge \bf  Thermal Spinning Brane Probes}

\vskip 1.5cm

\centerline{\large {\bf Jay Armas$\,^{1}$}, {\bf Niels A. Obers$\,^{2}$} and
{\bf Andreas Vigand  Pedersen$\,^{2}$} }

\vskip 1.0cm

\begin{center}
\sl $^1$ Albert Einstein Center for Fundamental Physics \\
Institute for Theoretical Physics, University of Bern \\
Sidlerstrasse 5, 3012 Bern, Switzerland
\vskip 0.4cm
\sl $^2$ The Niels Bohr Institute, Copenhagen University  \\
\sl  Blegdamsvej 17, DK-2100 Copenhagen \O , Denmark

\end{center}
\vskip 0.6cm

\centerline{\small\tt jay@itp.unibe.ch,  obers@nbi.dk, vigand@nbi.dk}

\vskip 1.3cm \centerline{\bf Abstract} \vskip 0.2cm \noindent

We construct and analyze  thermal spinning giant gravitons in type II/M-theory based on spherically wrapped black branes, using the
method of thermal probe branes originating from the blackfold approach. These solutions generalize in different directions 
recent work  in which the case of thermal (non-spinning) D3-brane giant gravitons was considered, and reveal a rich phase structure with various new properties.  First of all, we extend the construction to M-theory, by constructing thermal giant graviton solutions using spherically wrapped 
M2- and M5-branes.  More importantly, we switch on new quantum numbers, namely  internal spins on the sphere, which are not present in the usual extremal limit for which the brane world volume stress tensor is Lorentz invariant. 
We examine the effect of this new type of excitation and in particular analyze the physical quantities in various regimes, including that of
small temperatures as well as low/high spin.
As a byproduct we find new stationary dipole-charged black hole solutions in    $\ads_m \times S^n$  backgrounds of type II/M-theory. We finally show,  via a double scaling extremal limit, that our spinning thermal giant graviton solutions lead to a novel null-wave zero-temperature giant graviton solution with a BPS spectrum, which does not have an analogue in terms of the conventional weakly coupled world volume theory.

\end{titlepage}


\tableofcontents

\section{Introduction}

The use of brane probes in string/M-theory, notably the AdS/CFT correspondence, has been 
an important tool to uncover new physics and has generated a plethora of beautiful applications. 
In particular,  it has revealed novel features of supergravity backgrounds, phase transitions, stringy 
observables, non-perturbative aspects of field theories and dual manifestations of operators in CFTs.
Furthermore, a great deal has been learnt about the fundamentals of string/M-theory by studying the low energy
theories living on D/M-branes.

Recently a new method for thermal probe branes, based on the blackfold approach \cite{Emparan:2009cs},  has been developed and applied to various cases of interest 
\cite{Grignani:2010xm,Grignani:2011mr,Grignani:2012iw,Niarchos:2012pn,Niarchos:2012cy,Armas:2012bk,Niarchos:2013ia}. 
This has revealed a number of new qualitative and quantitative effects, as compared to the conventional method for probe branes in finite temperature backgrounds.  In this setting, Ref.~\cite{Armas:2012bk} found and analyzed thermal giant graviton solutions in type IIB string theory,
based on spherically wrapped black D3-branes. The purpose of this paper is to extend these results to a wider class of solutions, revealing a number of novel effects. In one direction, we extend the construction of  \cite{Armas:2012bk} to M-theory, by constructing thermal giant gravitons solutions using wrapped  M2- and M5-branes. In another direction, we generalize
the construction by switching on new quantum numbers, namely {\it internal spins} on the sphere, which are not present in the extremal limit due to the Lorentz invariance of the world volume stress tensor in that case. We examine the effect of this new type of excitation  and analyze the resulting phase structure and physical quantities in various regimes.
As a byproduct we note that by setting the angular velocity on the $S^1$ equal to zero, new stationary
blackfold solutions in  $\ads_m \times S^n$  backgrounds of type II/M-theory are found. 
Moreover, by applying a double scaling extremal limit to our thermal spinning giant graviton solutions, we find
a novel  null-wave giant graviton solution which exhibits a BPS spectrum and does not have a counterpart in the
usual weakly coupled world volume theory description. 

The physics of probe branes is conventionally examined using the weakly coupled description in terms of the D-brane (DBI) or M-brane world volume theories or, in the case of fundamental string probes, the Nambu-Goto action. However, as a consequence of ``open-closed string duality''%
\footnote{We use this terminology here loosely to denote a duality between world volume and space-time description.}
this weakly coupled (microscopic) world-volume picture has a complementary description on the strongly coupled (macroscopic)
bulk space-time side. Indeed, for supersymmetric configurations one can typically find an exact interpolation between these two sides, which has been at the heart of, for example, the microscopic counting of black hole entropy and the  AdS/CFT correspondence.

When considering the bending of supersymmetric brane configurations most work has been done by considering the world volume theory of a single brane in a given background. However, one expects that the corresponding brane profiles can also be obtained  from a gravity perspective by considering the back reaction of many branes on top of each other and solving the supergravity equations of motion using appropriate ans\"atze incorporating the symmetries.  A well-known example of this, relevant to the present paper,  is the relation between giant gravitons \cite{McGreevy:2000cw}
 and LLM geometries \cite{Lin:2004nb}, but more generally, this type of open-closed duality has been shown to extend beyond the AdS/CFT decoupling 
limit. For example, in Ref.~\cite{Lunin:2007mj}  the shapes of brane intersections were studied from the supergravity perspective
and found to be in perfect agreement with those found from the DBI action, the BIon solution of \cite{Callan:1997kz}
being the simplest example of this.  

A natural question is then whether one can extend these open-closed descriptions to the case of finite temperature 
(non-supersymmetric) configurations. This is first of all interesting in view of the fact that in many applications branes are used to
 probe finite temperature backgrounds. Furthermore, on the gravity side, branes become black when heated up, i.e. they develop 
 horizons,  so we may learn more about black hole physics. Finally, in the AdS/CFT context this provides us with nontrivial information
on thermal states in the dual field theories. On the world volume side this involves thermalizing the brane actions\footnote{See e.g. \cite{deBoer:2008gu}, where
the Nambu-Goto action was quantized in a finite temperature background.}
and subsequently finding non-trivial solutions, while on the space-time side one should find the corresponding gravity solutions of bent black branes. Since already at zero
temperature the latter leads to complicated (generally unsolvable) equations of motion, one approach is to treat the black branes
as finite temperature probes of the background.  This corresponds to the leading order blackfold method  
\cite{Emparan:2009cs}, which
thus provides us with a tool%
\footnote{Reviews include \cite{Emparan:2009zz} and  \cite{Camps:2012hw}  gives a more general derivation of the blackfold effective theory. See also Refs.~\cite{Emparan:2011hg,Caldarelli:2010xz,Grignani:2010xm} 
for the generalization of the blackfold approach to charged black branes.} 
to construct the finite temperature geometries in a perturbative expansion. 

This new method  to study thermal probe branes  has been used to study 
the thermalized version of the BIon system for the D3-brane \cite{Grignani:2010xm,Grignani:2011mr}, 
the gravity dual of the rectangular Wilson loop as described by an F-string ending on the boundary of $\ads_5\times S^5$ \cite{Grignani:2012iw}, the M2-M5 version of the BIon system \cite{Niarchos:2012pn,Niarchos:2012cy}, including 
a spinning M2-M5 ring intersection \cite{Niarchos:2013ia} and thermal giant gravitons  in type IIB string theory
\cite{Armas:2012bk}. 

In particular, Ref.~\cite{Armas:2012bk} analyzed what happens to the type IIB D3-brane giant graviton as one heats up the $\ads_5\times S^5$ background to non-zero temperature, requiring the D3-brane probe to thermalize with the background.
Several interesting new effects were found, including that the thermal giant graviton has a minimal possible value for the
angular momentum and correspondingly also a minimal possible radius of the $S^3$ on which the D3-brane is wrapped. Furthermore, the free energy of the thermal D3-brane giant graviton was computed in 
 the low temperature regime, which potentially can be compared to that of a thermal state on the
gauge theory side. A detailed analysis of the space of solutions and
stability of the thermal giant graviton was made and it was shown that, in parallel with the extremal case, there are two
available solutions for a given temperature and angular momentum, one stable and one unstable. 
The thermal giant graviton expanded in the $\ads_5$ part was also briefly examined in \cite{Armas:2012bk}.

The aim of the present paper is to include and analyze the effect of internal spin on the sphere on which the thermal
giant graviton is wrapped. In the process we will also generalize the results of \cite{Armas:2012bk} to M-theory, as we will
treat the D3, M2 and M5-brane cases in parallel. We will primarily focus on giant gravitons wrapping the sphere part of
the corresponding $\ads_m \times S^n$ backgrounds.  The possibility of adding internal spin is a new feature of thermal giant gravitons that is not present in the case of the standard extremal (supersymmetric) giant gravitons.  The reason is that
at zero temperature the world volume stress tensor of the giant graviton is locally Lorentz invariant, as can be seen directly from the D/M-brane
actions (for zero world volume gauge fields).  This means that the internal spin of the giant graviton is not  visible in the extremal limit. However turning on a temperature breaks the local Lorentz invariance of the world volume stress tensor and thus makes internal spin an important effect to consider.
Moreover, we find that it is possible to perform a non-trivial double scaling extremal limit, giving rise to a novel null-wave%
\footnote{Null-waves were first considered in the blackfold context in Ref.~\cite{Emparan:2011hg}. 
Furthermore, a null-wave on the M2-M5 brane intersection was recently considered in Ref.~\cite{Niarchos:2013ia} using blackfold methods.}
giant graviton with BPS spectrum.

A short outline and main results of the paper are as follows: 

We start in Sec.~\ref{sec:setup} by setting up the problem and deriving the blackfold action and resulting equations of motion
describing thermal spinning giant gravitons obtained by wrapping $(n-2)$-branes on an $S^{n-2}$ sphere
in the sphere-part of $\ads_m \times S^n$, where the cases of interest are $(m,n)=\left\{ (5,5) ;(4,7);(7,4)\right\}$.
  Our discussion will be succinct, and we refer to \cite{Armas:2012bk} for more details. 
The giant graviton is rotating on an $S^1$ of the $S^n$, and the new aspect we consider here is that it is 
simultaneously spinning on the $S^{n-2}$. This is only possible for odd $n$, so the M2-brane case included in this paper 
is by construction non-spinning, while for the D3 and M5-brane we focus here on the maximally symmetric case with equal
spins in each of the $(n-1)/2$ Cartan directions. The solution of the equations of motion is  presented, for given temperature $T$, number of branes $N_{(n-2)}$ and internal spin $\CS$, and expressions for the various physical quantities are given. The picture that emerges is that there are two branches of solutions, a lower and upper branch, as was seen also in \cite{Armas:2012bk}. However, in the presence of internal spin each of these two branches splits up further into two branches, a low spin and high
spin branch.  We also consider the regime of validity of our approach, in which the $(n-2)$-branes are treated as probes with locally approximately flat world volume. This leads to the requirement that $ 1 \ll N_{(n-2)} \ll N^{\frac{m-1}{n-1}}$, where $N$ is the quantized flux of the background. The issue of the Hawking-Page transition is also addressed. 

Sec.~\ref{sec:thermal} is devoted to examining the solution space of the thermal spinning giant gravitons in further detail
(this is supplemented by App.~\ref{app:reparameterization} which  discusses various other properties of the solution space). 
The main features of the solution space are exhibited by plotting angular velocities for a representative value of the temperature.
It is shown that the regimes of low and high spin have distinct properties. For low spin, the thermodynamics receives quadratic spin corrections which are subleading to the thermal corrections from the non-zero temperature. On the other hand, in the high spin regime (which is bounded by a given value of maximal internal spin) the physics is dominated by the effects of internal spin, and we present perturbative expressions for the physical quantities in that regime.  We furthermore perform a low temperature
expansion, obtaining first the free energies for the non-spinning thermal giant graviton (see Eq.~\eqref{freeenergy}). It is interesting
to observe that the leading thermal contribution to $F-J/L$ (with $F$ the free energy
and $J$ the angular momentum on $S^1$) in each case is proportional to the free energy of the field theory living on the giant graviton brane, independent of $J$ and $N$. We furthermore consider the cases of low temperature with in addition  low and high intrinsic spin respectively.
Finally, as a byproduct of our analysis we note that, once intrinsic spin is introduced, one can also solve the equations
of motion for the case $\Omega=0$, i.e. no rotation on the $S^1$ and only intrinsic spin. Hence the resulting solution is stationary%
\footnote{For $J \neq 0$ the solution is ``quasi-stationary", see  \cite{Armas:2012bk} for a detailed discussion.}
so that we find, in the blackfold limit,  a novel stationary black hole solution in $\ads_m \times S^n$, in analogy with stationary odd-sphere blackfold solutions in asymptotically flat space 
\cite{Emparan:2007wm,Emparan:2009vd,Caldarelli:2010xz,Emparan:2011hg} and $\ads$ space \cite{Caldarelli:2008pz}. For the D3/M5-brane case these solutions
have horizon topology $S^5 \times S^3$/$S^4 \times S^5$ respectively, and the solution carries brane dipole charge. These are the first examples of such novel black holes space times in $\ads_5 \times S^5$ and $\ads_4 \times S^7$
respectively. 

In Sec.~\ref{sec:nullwave} we find a zero temperature excitation of the usual extremal giant graviton by taking a double scaling
extremal limit of our thermal spinning D3/M5-brane giant graviton solutions.  The resulting solution is described by
a null-wave on the giant graviton world volume, and we call this a null-wave giant graviton. The spectrum of the solution
is computed and we show in particular that the lower branch satisfies $\textbf{E}=\textbf{J}+\boldsymbol{\mathcal{S}}$
with $\boldsymbol{\mathcal{S}}$ the total spin carried by the null-wave on the internal sphere. We take this as a strong indication that these solutions are $\frac{1}{8}$-BPS for the D3-brane case and $\frac{1}{16}$-BPS
for the M5-brane case. The spectrum of the upper branch in the null-wave limit is also obtained. 
Furthermore, a stability analysis of the lower and upper branch is presented, and in particular it is shown that the lower branch is stable as expected for a BPS solution. The null-wave giant gravitons are new and do not have a counterpart in the standard weakly coupled brane world volume theories. Using the blackfold action as a starting point, we also present an action that describes these null-wave giant gravitons. 
This action is then used to construct in addition null-wave giant gravitons expanded into the $\ads$ factor of the background. We observe the same BPS spectra as for the null-wave giant gravitons expanded on the sphere part.

\section{Setup using thermal probe method \label{sec:setup} }

In this section we discuss the setup that we employ to obtain thermal spinning giant gravitons.  The method uses the blackfold approach
\cite{Emparan:2009cs,Emparan:2011hg,Caldarelli:2010xz,Grignani:2010xm}
to thermal probe branes in string theory, and parallels in particular the case of thermal (non-spinning) D3-brane giant gravitons discussed in Ref.~\cite{Armas:2012bk}, to which we refer the reader for further details and choice of notation.  This is generalized here to include i)  thermal M2, M5-brane giant gravitons and ii) internal spin. 
As we will see the latter can only be consistently turned on for odd branes (i.e. the D3 and M5-brane case).
 Beyond the setup and the resulting blackfold equation of motion, this section presents the corresponding thermal spinning giant gravitons solutions, the regime of validity and the extremal limit.

\subsection{Blackfold action and equation \label{sec:action} }

Our aim is to study giant graviton solutions of type II string theory and M-theory as the $\ads_m\times S^n$ background is heated up to finite temperature,
treating the giant gravitons as probes of these backgrounds, but heating them up to the same (finite) temperature. This is done by going to the supergravity regime and replacing the thermal probe branes by an effective description in terms of their stress tensor and charge current.

We focus here on the conformal cases, namely we will be considering D3-branes in the type IIB supergravity background
and  M5- and M2-branes in the $D=11$ supergravity backgrounds  of the form  $\ads_m\times S^n$,  with 
$(m,n)=\left\{ (5,5);(4,7);(7,4)\right\}$. Note that $n$ and $m$ are related by  $n=(3m-5)/(m-3)$ but  for ease of notation we will keep 
thse symbols separately below. We restrict our attention in this paper primarily to the corresponding $(n-2)$-branes wrapped on an $S^{n-2}$  inside
the $S^n$-sphere of the background (except in Sec.~\ref{sec:nwads}). The motion of this thermal probe brane (blackfold) of topology $S^{n-2}$ 
is supported by the background gauge field on the $S^n$.  The analysis is readily generalized to wrapping the branes on spheres inside the $\ads$ factor, as was discussed for thermal (non-spinning) D3-brane giant gravitons in \cite{Armas:2012bk}. 

Our first input to set up the problem is the stress tensor and charge current of the black $(n-2)$-brane probes. To leading order in the  blackfold approximation 
the stress tensor is that of a $(n-1)$-dimensional perfect fluid tensor $T_{ab} = (\epsilon + P) u_a u_b + P \gamma_{ab}$ where
$\sigma^a=\tau, \sigma^1 \ldots  ,\sigma^{n-2}$ label the world volume coordinates,  $u_a$ is  the $(n-1)$-velocity and $\gamma_{ab}$ the induced
metric on the brane. Furthermore, the energy, pressure, entropy density and local temperature are given by 
\begin{equation}\label{thermoquantities}
\epsilon = \mathcal T s - P  \spa P = -\mathcal{G}\left(1+ (m-1) \sinh^2 \alpha\right) \spa \mathcal T s=(m-1)  \mathcal{G}
\spa \mathcal T = \frac{m-1}{4 \pi r_0 \cosh \alpha }  \  , 
\end{equation}
where we have defined
\begin{equation}
\label{Gdef}
 \mathcal G \equiv  
\frac{\Omega_{(m)}}{16 \pi G}   r_0^{m-1}  \  , 
\end{equation}
with  $\Omega_{(m)}$ the volume of the unit $m$-sphere. 
The parameters of the black $(n-2)$-brane stress tensor and thermodynamics are thus $r_0$,~$\alpha$ and the codimension
of the brane $m+1$.  Note that we can replace Newton's constant $G$ in terms of the tension $T_{(n-2)} = ( (2 \pi)^{n-2} l_p^{n-1} )^{-1}$  of the $(n-2)$-brane using the relation%
\footnote{Note that for this one uses $16 \pi G = (2\pi)^{m+n-3} l_p^{m+n-2} $, where we recall that for the IIB string theory
case $l_p^8 = g_s^2 l_s^8$.}
\begin{equation}
\label{eq:tension}
T_{(n-2)}=\frac{N}{\Omega_{(n-2)}L^{n-1}} \  , 
\end{equation}
where $N$ and $L$ are the magnitude of the flux and the radius of the sphere part of $\ads_m\times S^n$ respectively.  
The black $(n-2)$-brane furthermore has the $(n-1)$-form charge current 
$J_{(n-1)} = Q_{(n-2)} d\tau \wedge d\sigma^1 \wedge  \ldots \wedge d \sigma^{n-2} $ 
where  $Q_{(n-2)}$ is the charge density
\begin{equation}
\label{branecurrent} 
{Q}_{(n-2)} =(m-1) \mathcal G \sinh \alpha \cosh \alpha = N_{(n-2)} T_{(n-2)} \ , 
\end{equation}
and $N_{(n-2)}$ the number of probe black $(n-2)$-branes. Note that current conservation on the world volume implies that
$Q_{(n-2)}$ is constant. 

Turning to the background and the embedding of the probe brane, we write the metric on $S^n$ as
\begin{equation}
\label{spherecoord}
d\Omega_{(n)}^2=L^2\left(\text{d}\theta^2+\cos^2\theta \text{d}\phi^2 + \sin^2\theta d\Omega_{(n-2)}^2 \right) \  . 
\end{equation}
The giant graviton spatial world volume is spanned by the $S^{n-2}$ and it moves around the $S^1\subset S^n$ described by the coordinate $\phi$ with angular velocity $\dot \phi\equiv  \beta_n \Omega$, $\beta_n =(-1)^{D-n-1}$\footnote{The choice of sign $\beta_n$ is introduced for convenience to simplify the formulae below, treating the D3, M2 and M5-branes uniformly.
Alternatively, one can take a plus sign for all cases, and reverse the sign of the M5-brane charge, turning it into an anti-M5-brane.}.
The size $r$ of the giant graviton and the distance to
the equator of the $S^n$ is described by the $\theta$ coordinate, $r\equiv L \sin\theta$.  As mentioned above, in addition to generalizing the D3-brane giant graviton to M-branes, the main objective of this paper is to examine the effects of intrinsic spin. 
To incorporate this, we write the spatial part of the induced metric on the brane as $d\Omega_{(n-2)}^2=\sum_{i=1}^{[n/2]}\text{d}\mu_i^2+\sum_{j=1}^{[(n-1)/2]} \mu_j^2\text{d}\phi_j^2$ subject to the condition $\sum_{i=1}^{[n/2]} \mu_i^2=1$. Then,  instead of considering the 
effective fluid at rest $u^a\partial_a \sim \partial_\tau$, we will now consider the following fluid velocity
\begin{equation}
u^a=\frac{\bk^a}{\bk}~~, \quad \bk^a \partial_a=\partial_\tau+\omega\sum_{i=1}^{[(n-1)/2]}  \partial_{\phi_i}    \  ~,
\end{equation}
where we have defined $\bk\equiv|-\gamma_{ab}\bk^{a}\bk^{b}|^{\frac{1}{2}}$. We thus take the maximally symmetric
situation with equal angular velocities in each of the Cartan directions of the $S^{n-2}$, and, for reasons explained below we will
assume that  $n$ is odd.  We then have the norms 
\begin{equation}
\label{kFnorm}
 \bk^2= |k_{\text{w.v.}}|^2 - {\CW}^2  \spa \CW \equiv \omega r   \spa 
 |k_{\text{w.v.}}|^2 \equiv |\partial_\tau|^2=1- (\Omega L)^2 +\CV^2 \spa \CV \equiv \Omega r  \  . 
\end{equation}
Note that for $n$ even the first expression above would depend on one of the direction cosines $\mu_{n/2}$, leading to a Killing vector with a norm that is angular  dependent. In analogy with the neutral blackfold solutions considered in Ref.~\cite{Emparan:2009vd}, this will lead to an inconsistency in the equation of motion.
Thus we can only consistently switch on internal spin for the D3 and M5-brane, where the branes wrap odd-spheres. The results below still hold for the M2-brane provided one sets the internal angular velocity   $\omega$ to zero.  

We will also need the background gauge field which in terms of the coordinates defined in \eqref{spherecoord}  takes the form
\begin{equation}
A_{[n-1]} = ( L \sin \theta)^{n-1} d\phi \wedge d \Omega_{(n-2)}  =
r^{n-1} d\phi \wedge d \Omega_{(n-2)} \  . 
\end{equation}
Given the embedding described above, pulling back the gauge form to the world volume gives a factor $\Omega$, the angular velocity of the giant graviton on the $S^1$.

\subsection*{Thermal giant graviton equation of motion}

We are now ready to derive the equation of motion (EOM) for the spinning thermal giant graviton. Here we derive the equation directly from the blackfold world volume action (for a derivation based on the blackfold extrinsic equation see  App.~C of \cite{Armas:2012bk}). The action takes the form 
\begin{equation} \label{action}
I=\int_{\mathbb{R}\times S^{(n-2)}}\left\{*P+Q_{(n-2)}\mathbb{P}\left[A_{[n-1]}\right]\right\}  \  , 
\end{equation}
where  $\mathbb{R}$ denotes time, $\mathbb{P}\left[A_{[n-1]}\right]$ is the pull-back of the background gauge field $A_{[n-1]}$ to the world volume and $Q_{(n-2)}=N_{(n-2)} T_{(n-2)} $ is the total charge of the giant graviton (see also  \eqref{branecurrent}). 
We also remark that since the $(n-2)$-brane is expanded on the $(n-2)$-sphere the local temperature has a redshift as compared to the global temperature $T$ of the background space-time that we are probing, i.e. $\mathcal T =  T /\bk$ .

Using the embedding given above, and employing the $\text{SO}(n-1)$ symmetry 
of the configuration, the action takes the form 
\begin{equation}
\label{actionGG}
\beta I_E   = -\Omega_{(n-2)}  r^{n-2}\left(|k_{\text{w.v.}}|P+ r\Omega Q_{(n-2)}\right)  \  , 
\end{equation}
where we have gone to Euclidean space and the factor $\beta=1/T$ results from the integration over Euclidean time. 
The equation of motion is obtained by varying the action keeping fixed $(T,\Omega,\omega)$ and $Q_{(n-2)}$. 
Using the definitions in \eqref{kFnorm} and the identity $ \delta_r \log P=-\mathcal R_1 \delta_r \log \bk $
we find after some algebra   the EOM in the form
\begin{equation}\label{exteq}
(n-2)\left(\mathbf{k}^2+\mathcal{W}^2\right)+\mathcal{V}^2+\frac{\mathbf{k}^2+\mathcal{W}^2}{\mathbf{k}^2}\mathcal{R}_1\left(\mathcal{W}^2-\mathcal{V}^2\right)+ (n-1) \mathcal{V} \sqrt{\mathbf{k}^2+\mathcal{W}^2} \mathcal R_2=0 \  , 
\end{equation}
where we have introduced the two ratios \cite{Armas:2012bk} 
\begin{equation}
\label{R12}
\mathcal R_1\equiv \frac{\mathcal T s}{P}=\frac{1-m}{1+(m-1)\sinh^2\alpha}, \quad \text{and} \quad 
\mathcal R_2\equiv \frac{Q_{(n-2)} }{P}=\frac{(1-m)\sinh \alpha \cosh \alpha}{1+(m-1)\sinh^2\alpha} \ . 
\end{equation}

\subsubsection*{Conserved quantities}
Given a solution of the EOM \eqref{exteq}, the configuration has a number of conserved quantities. 
For use below, we present the (off-shell) expressions of these conserved quantities, which follow from the general results
for blackfolds in flux backgrounds, derived in \cite{Armas:2012bk}. These are given by 
\begin{equation} \label{conservedQ}
E=\frac{\Omega_{(n-2)} r^{n-2}}{|k_{\text{w.v.}}|\bk^2}\left[\epsilon |k_{\text{w.v.}}|^2+P\left(|k_{\text{w.v.}}|^2-\bk^2\right) \right] \spa 
S= \frac{1}{T}  (m-1)\Omega_{(n-2)}\mathcal G   |k_{\text{w.v.}}|  r^{n-2}  \ , 
\end{equation}
\begin{equation} 
\label{conservedJS}
J=E\Omega\rho^2+ Q_{(n-2)}\Omega_{(n-2)}r^{n-1}
\spa 
\mathcal S = 
\frac{ \Omega_{(n-2)} \mathcal{G}\omega r^{n}|k_{\text{w.v.}}|}{\bk^2} \ , 
\end{equation}
where $E$ is the energy, $S$ the entropy, 
$J$ the angular momentum along the $S^1 \subset S^{n}$,  and 
$\mathcal S_i  = \frac{2}{n-1} \mathcal S$, $i=1,\ldots (n-1)/2$  the intrisinc angular momenta on $S^{n-2}$.  Here and in the following we
have also introduced $\rho = \sqrt{L^2 - r^2} $ and we remind the reader that $\epsilon$, $P$, ${\cal{G}}$ are defined in \eqref{thermoquantities}, 
\eqref{Gdef}.  Note that, in accord with the results of App.~B in \cite{Armas:2012bk}
one can check that the Euclidean action in  \eqref{actionGG} satisfies $\beta I_E = F_{\rm G} = E - T S - \Omega J - \omega \mathcal S$. 
The equations of motion are therefore equivalent to requiring the first law of thermodynamics.

\subsection{Solution space and thermodynamics \label{sec:solution} }

We now describe the solution space of the EOM \eqref{exteq}. 
We work in the ensemble with given temperature $T$, fixed charge (number of $(n-2)$-branes) $Q_{(n-2)}$ 
and intrinsic spin $\mathcal S$. As in \cite{Armas:2012bk} we will use the norm of the fluid killing vector $\mathbf k$ to (formally) parameterize the solution space as follows. For a given $\mathbf k, \mathcal R_1, \mathcal R_2$ and $\mathcal W$
 we can solve the EOM \eqref{exteq} for $\CV$ since it is a simple quadratic equation,
\begin{equation}\label{eqq1}
\mathcal V_\pm  (\bk, \CW) =\frac{1}{2}\frac{ (n-1)\mathcal R_2\sqrt{\mathbf{k}^2+\mathcal W^2}\mp\sqrt{\mathcal D_{\mathcal{W}}^{(n)}}}{(\mathcal{R}_1-1)\mathbf{k}^2+\mathcal R_1\mathcal W^2}\mathbf{k}^2 \ , 
\end{equation}
with 
\begin{equation}\label{eqq2}
\mathcal D_{\mathcal{W}}^{(n)}=\left(\mathbf{k}^2+\mathcal W^2\right)\left(4\left(n-2+\frac{\mathcal R_1}{\mathbf{k}^2}\mathcal W^2\right)\left(\mathcal R_1-1+\frac{\mathcal R_1}{\mathbf{k}^2}\mathcal W^2\right)+(n-1)^2\mathcal R_2^2\right) \ . 
\end{equation}
We will refer to the two solution branches as the lower ($-$) and upper ($+$) branch respectively. At the end of this section, we will
show that for zero intrinsic spin and zero temperature the lower branch reduces to the standard $\frac{1}{2}$-BPS giant graviton, while
in that limit the upper branch is another extremal solution that has not received much attention in the literature\footnote{See also \cite{Armas:2012bk} where the extremal limit of the lower branch for the D3-brane case was discussed in detail.}. 
Using \eqref{kFnorm} we can now find the expression for respectively 
$\hat r \equiv r/L$, $\hat \Omega \equiv \Omega L$ and $\hat \omega \equiv \omega L $. One finds
\begin{equation}\label{rOmegaomega}
\hat r (\bk, \CW) =\frac{\mathcal V}{\sqrt{1+\mathcal V^2-\mathcal W^2-\mathbf{k}^2}}, \quad  \hat\Omega (\bk, \CW) =\frac{\mathcal{V}}{\hat r}, \quad
 \hat \omega (\bk, \CW) =\frac{\mathcal{W}}{\hat r}  \ , 
\end{equation}
where $\CV (\bk, \CW) $ is given by  \eqref{eqq1}.  
Now for a given value of $\mathbf k$ we can (explicitly) work out the values of $\mathcal R_1\equiv \mathcal R_1(\mathbf{k})$ and $\mathcal R_2(\mathbf{k})$ and (implicitly) the value of $\mathcal W\equiv \mathcal W (\mathbf{k})$ by the requirement that $Q_{(n-2)}, T$ and $\mathcal S$ are kept fixed. This will be explained in the following. 

First of all, following \cite{Armas:2012bk}, we can determine the value of $\mathcal R_1$ and $\mathcal R_2$ (see
\eqref{R12}) 
for a given $\mathbf k$. To this end, we introduce the parameter $\phi\equiv 1/\cosh^2\alpha$.  The charge conservation equation \eqref{branecurrent} can then be rewritten as
\begin{equation}\label{eq2}
\phi^{m-2}-\phi^{m-3}+\frac{(m-3)^{m-3}}{(m-2)^{m-2}}\sin^2\delta=0   \ , 
\end{equation}
with%
\footnote{Note that $\hT$ here is slightly differently defined as compared to \cite{Armas:2012bk}.}
\begin{equation}
\label{sindelta}
\sin\delta=\left(\frac{\hT }{\bk}\right)^{m-1} \spa \hT \equiv \frac{T}{T_{\rm stat}} \spa T_{\rm stat}^{m-1} = \frac{1}{Q_{(n-2)} G}
\frac{(m-1)^m \Omega_{(m)}}{4(4\pi)^m }
\sqrt{\frac{(m-3)^{m-3}}{(m-2)^{m-2}}} \ . 
\end{equation}
The equation \eqref{eq2} is a polynomial of degree $m-2$ whose solution we will denote by $\phi(\mathbf{k})$ (for simplicity of notation we suppress the $\hT$ dependence in all expressions below). 
For $m=4$ (M5-brane on $S^7$), it becomes a simple quadratic equation with solution
\begin{equation}
(m,n)=(4,7): \qquad  \phi(\mathbf k)=\sin^2\left(\frac{\delta(\mathbf k)}{2}\right)  \  .
\end{equation}
In the case of $m=5$ (D3-brane on $S^5$), the equation becomes cubic with solution \cite{Armas:2012bk}
\begin{equation}
(m,n)=(5,5) \qquad \phi(\mathbf k)=\frac{2}{3}\frac{\sin\left(\delta(\mathbf k)\right)}{\sqrt{3}\cos\left(\delta(\mathbf k)/3\right)-\sin\left(\delta(\mathbf k)/3\right)} \ . 
\end{equation} 
It is not possible to write down an analytical expression for $m=7$ (M2-brane on $S^4$)
but  $\phi(\mathbf k)$ can be obtained numerically. 

The second parameter $\mathcal W$ is determined by the (fixed) intrinsic spin $\mathcal S$. Rewriting $\mathcal S$ is straightforward using
the expression in \eqref{conservedJS}. We have 
\begin{equation}\label{eq3}
\mathcal{S} (\bk, \CW) = L Q_{(n-2)}  \Omega_{(n-2)}\frac{\phi(\mathbf k) \mathcal W \sqrt{\mathbf k^2+\mathcal W^2}}{\mathbf k^2 \sqrt{1-\phi(\mathbf k)}} \hat r\left(\bk, \mathcal W\right)^{n-1} \  , 
\end{equation}
where we recall that $\hat r$ is given in \eqref{rOmegaomega}.This equation does not in general have an analytical solution but it is a simple algebraic equation in one variable $\mathcal W$ and its solution is again in principle easy to obtain numerically. We 
denote the solution by $\mathcal W(\mathbf k)$.

 The equations \eqref{eqq1}-\eqref{sindelta} and \eqref{eq3}
 formally parameterize the solution in terms of $\mathbf k$ for given $\hT$, $Q_{(n-2)}$ and $\CS$. 

\subsubsection*{Range of $\mathbf{k}$}

  Finally, we need to address the range of $\bk$.  
First of all we note that $\bk$ necessarily lies in the range $\hat T \leq \bk \leq 1$, where the lower bound follows from
\eqref{sindelta} and the upper bound from the geometric relation $r \leq L$. 
However, this is only a necessary condition and the form of the solution, notably positivity of the discriminant in \eqref{eqq2},
leads to further restrictions.  In particular, for the non-spinning giant graviton ($\CS=0$) this leads to the  
 restricted range $\tilde{T} = T/T_{\rm max} \leq \mathbf k \leq 1$.  Here $T_{\rm max}$ is the maximum temperature that
 the solution can have in that case (see App.~\ref{app:detailsmaxtemp}), and we note that $\hat T < \tilde T$ because 
 $T_{\rm stat} > T_{\rm max} $.  More generally, as soon as we turn on spin one finds that the range of possible $\bk$ values
  becomes more intricate but can be computed in principle for given $\hat T$, $\CS$. 

As an illustration we given some details on the range of $\bk$ in App.~\ref{app:detailsrange}, while
we also refer the reader to Sec.~\ref{sec:thermal}, where we will plot the solution branches for a representative value of $\hT$. 
 This indicates that  $\bk$  goes from 1 (low spin regime) to $\hT$ (for which the maximum spin is obtained)
 and  a small interval of $\bk$'s which is excluded by the EOM.  As a consequence, we see that each of the lower and upper branches, branch up further into two branches, a low spin and high spin branch.

\subsubsection*{Physical quantities}

Given a spinning giant graviton solution, we can write down the on-shell physical quantities using the expressions
in \eqref{conservedQ}, \eqref{conservedJS}.  As in \cite{Armas:2012bk} we define rescaled dimensionless energy, entropy,
and angular momenta by 
\begin{equation}
\mathbf{E} \equiv \frac{EL}{N N_{(n-2)}}, \quad   \mathbf{S}\equiv \frac{S T_{\text{stat}} }{N N_{(n-2)}    }, \quad
\mathbf{J}\equiv \frac{J}{N N_{(n-2)}}, \quad
\boldsymbol{\mathcal{S}} \equiv \frac{\mathcal S}{N N_{(n-2)}} \ , 
\end{equation}
and use the dimensionless ratios $ \hat r\equiv r/L$  $\hat \rho \equiv \rho/L$.  
Notice that $\mathbf J$  (respectively $\boldsymbol{\mathcal{S}}$) is the ratio between orbital (respectively internal) angular momentum and the orbital angular momentum of the maximal
size giant graviton at $r=L$. 
We then record the expressions of $\mathbf E$,  $ \mathbf S$, $\mathbf J$ and $\boldsymbol{\mathcal{S}}$ in terms of $\mathbf k$,  $\mathcal{W}$, $\phi (\bk)$ and $ \hat r( \bk, \CW)$
\begin{equation}
\label{EJ}
\mathbf E=\frac{1}{\sqrt{\mathbf k^2+\mathcal W^2}}\frac{1+\frac{\phi}{\mathbf{k}^2}\mathcal W^2+\frac{\phi}{m-1}}{\sqrt{1-\phi}}\hat r^{n-2}, \quad 
 \mathbf S= \frac{1}{\hat T} \frac{\phi}{\sqrt{1-\phi}}\sqrt{\mathbf k^2+\mathcal W^2}\hat{r}^{n-2} \  , 
\end{equation}
\begin{equation}
\label{SS}
\mathbf J=\boldsymbol{E}\hat{\rho}\sqrt{1-\mathcal W^2-\mathbf k^2}+ \hat{r}^{n-1} \ , 
\quad
\boldsymbol{\mathcal{S}}=\frac{\phi \mathcal W \sqrt{\mathbf k^2+\mathcal W^2}}{
\mathbf k^2 \sqrt{1-\phi}}\hat r^{n-1} \ . 
\end{equation}
 The expression for $\boldsymbol{\mathcal{S}}$ suggests that maximum intrinsic spin is attained for
$\bk = \hat T$, which is confirmed by the analysis in the next section. 

\subsubsection*{Validity of the approach}

We also address the validity of the (leading order) blackfold approach in which the $(n-2)$-brane is treated in the probe
approximation. For the probe approximation to be valid for our supergravity black $(n-2)$-brane probe we must require the transverse length scale $r_s$ of the probe to satisfy the conditions that 
$r_s$ is much smaller than any of the scales $ r_{\rm int}$, $r_{\rm ext}$ and  $L$, 
where $r_{\rm  int}$ and $r_{\rm ext}$ are the length scales associated with the intrinsic and extrinsic curvature of the embedding of the brane, respectively, and $L$ is the length scale of the $\ads_m \times S^n$ background. 
A detailed analysis leads to the (sufficient) requirement  
\beq
\label{validity} 
1 \ll N_{\rm D3} \ll N \ll \lambda N_{\rm D3} \ , \quad 1 \ll N_{\rm M5}^2 \ll N \ , \quad 1 \ll N_{\rm M2}  \ll N^2  \ . 
\eeq
We note that the upper bounds $ N_{(n-2)} \ll N^{\frac{m-1}{n-1}}$ follow from setting $r=L$ in  the necessary requirement 
$ N_{(n-2)} \ll  N^{\frac{m-1}{n-1}} (r/L)^{m-1} $. It is interesting to observe that the last two conditions (for the M-branes) can be
rewritten as $\lambda_M \ll 1$ and $\lambda_M \gg 1$ respectively, in  terms of the 't Hooft like coupling
$\lambda_M= N_5^2/N_2$ that was identified  in Ref.~\cite{Niarchos:2012cy} in the context of the self-dual string soliton
of the M5-brane theory. Here we use the fact that for the M5-brane case our $N $ is the parameter of the M2-brane theory
and for the M2-brane case $N$ is the parameter of the M5-brane theory.

It is also important to examine how these bounds relate to the Hawking-Page temperature $T_{\rm HP} \sim 1/L$, above
which the $\ads$ black hole background will become dominant over the hot $\ads$ space-time background
considered in this paper. Using the results for the maximal temperature collected in App.~\ref{app:detailsmaxtemp}
we have first of all in the case of zero intrinsic spin that
\beq
\frac{T_{\rm max}  }{T_{\rm HP}} \sim \frac{N^\frac{1}{n-1} }{N_{(n-2)}^\frac{1}{m-1} }   \gg 1 \ , 
\eeq
where we used \eqref{validity} in the last step. We thus see that  in the regime where the probe blackfold approximation
is valid, the maximum temperature of the solution is far above the Hawking-Page temperature. 
As a consequence this maximum temperature is not physical in the sense that before reaching it one should
change the background to the $\ads$ black hole, and hence our solution is most relevant for small temperatures
($\hT \ll 1$).  We also remark that when the intrinsic spin is turned on the maximum temperature decreases.

\subsection{Extremal limit \label{sec:extremal} }

To make contact with the standard zero-temperature giant graviton we consider here the extremal limit. This is obtained by setting $\phi=0$ so $\mathcal R_1=0$ and $\mathcal R_2=-1$.  
Since $\boldsymbol{\mathcal{S}}=0$ for all $\mathcal W$, we expect $\mathcal W$ to drop out of the
problem\footnote{Another extremal limit, involving a double scaling, will be considered in Sec.~\ref{sec:nullwave}.}.
Indeed, we do not expect to be able to see intrinsic rotation in the extremal limit, due to Lorentz invariance of the world volume stress tensor. In further detail, we obtain from the solution \eqref{eqq1} by setting  $\mathcal R_1=0$ and $\mathcal R_2=-1$
that
\begin{equation}
\mathcal V_-=\sqrt{\mathbf{k}^2+\mathcal W^2}=|k_{\text{w.v.}}|, \quad \mathcal V_+=(n-2)\sqrt{\mathbf{k}^2+\mathcal W^2}=(n-2)|k_{\text{w.v.}}|  \ . 
\end{equation}
Using that $|k_{\text{w.v.}}|=1-\hat \Omega^2(\hat r) \hat r^2, \ \mathcal V =\hat \Omega(\hat r) \hat r $, it is then easy to parameterize the angular velocity $\Omega$ in terms of the size of the giant graviton $\hat r $ 
\begin{equation}\label{extremalsolution}
\hat{\bar \Omega}_-(r)=1 , \quad \hat{ \bar \Omega}_+(r)=\frac{n-2}{\sqrt{(n-2)^2- (n-1)(n-3) \hat r^2}} \ , 
\end{equation}
and we verify that the results are independent of $\mathcal W$. Here the lower branch is the standard $\frac{1}{2}$-BPS giant graviton
while the upper branch (see also \cite{Armas:2012bk}) is a second extremal giant graviton branch. Our thermal giant graviton
branches thus correspond to heating up (and spinning up) these extremal solutions. 
The corresponding extremal results for the giant graviton on $\ads$ are obtained by the transformation
$\hat{\bar \Omega}_\pm (\ads) =  [ \hat{\bar \Omega}_\pm (\hat r \rightarrow i \hat r) ]^{-1}$. 

It is straightforward to compute the energy and angular momentum of the extremal solutions using \eqref{EJ} and the above.
For the lower branch we find 
\begin{equation}
\label{EandJextramalminus}
\mathbf E_-=\hat{r}^{n-3}, \quad \mathbf J_-=\hat{r}^{n-3} \ , 
\end{equation}
while the upper branch has 
\begin{equation}
\label{EandJextramalplus}
\mathbf E_+=\hat{r}^{n-3}\sqrt{\left(n-2\right)^2-\left(n-3\right)\left(n-1\right)\hat r^2}, \quad \mathbf J_+=\hat{r}^{n-3}\left(n-2+(n-3)\hat r^2\right) \ , 
\end{equation}
In Ref.~\cite{Armas:2012bk} the stability of both branches was examined in detail for the D3-brane case, which easily
generalizes in the obvious way to the M2- and M5-brane cases considered here. 

A point worth emphasizing is that the extremal solutions discussed here are in fact supergravity solutions: they represent 
backgrounds with probe supergravity branes in them, computed to leading order in the blackfold approach.
That these solutions directly map onto the D/M-brane giant graviton world volume solutions is a consequence of supersymmetry
(extremality). In this connection, note that the extremal limit of the blackfold world volume action \eqref{actionGG} is the D/M-brane
world volume action multiplied by a factor of $N_{(n-2)}$ the number of probe branes, which in the blackfold approach is very large. 

\section{Thermal spinning giant graviton \label{sec:thermal} }

In this section we examine the physics of the thermal and internally spinning version of the giant graviton configuration consisting of an $(n-2)$-brane wrapped
on an $(n-2)$-sphere moving on the $n$-sphere of $\ads_m \times S^n$. We will start by elucidating some of the main features of the solution space obtained from the EOM \eqref{exteq}.
\subsection{Main features of solution space}\label{sec:maxspin}
From the point of view of the dual field theory, the most interesting giant graviton configuration is the one close to maximal size, $r \simeq L$. In this case the dual operator (on the lower branch) is known. In the extremal case for $r=L$ there are two solutions to the EOM, namely $\hat{\overline{\Omega}}=1$ and $\hat{\overline{\Omega}}=n-2$ corresponding to the end points of the lower and upper branch, respectively (cf. Eq.~\eqref{extremalsolution}). In this section we examine the configuration space at $r=L$ when turning on temperature and intrisic spin. We mention that in principle it is possible to numerically do a similar analysis for any $r>0$, however, this is not particularly illuminating and such an analysis has thus been omitted. We expect the general features of the results below to hold for any $r$.  

\begin{figure}[b!]
\centering
\begin{subfigure}{.5\textwidth}
  \centering
  \includegraphics[width=.8\linewidth]{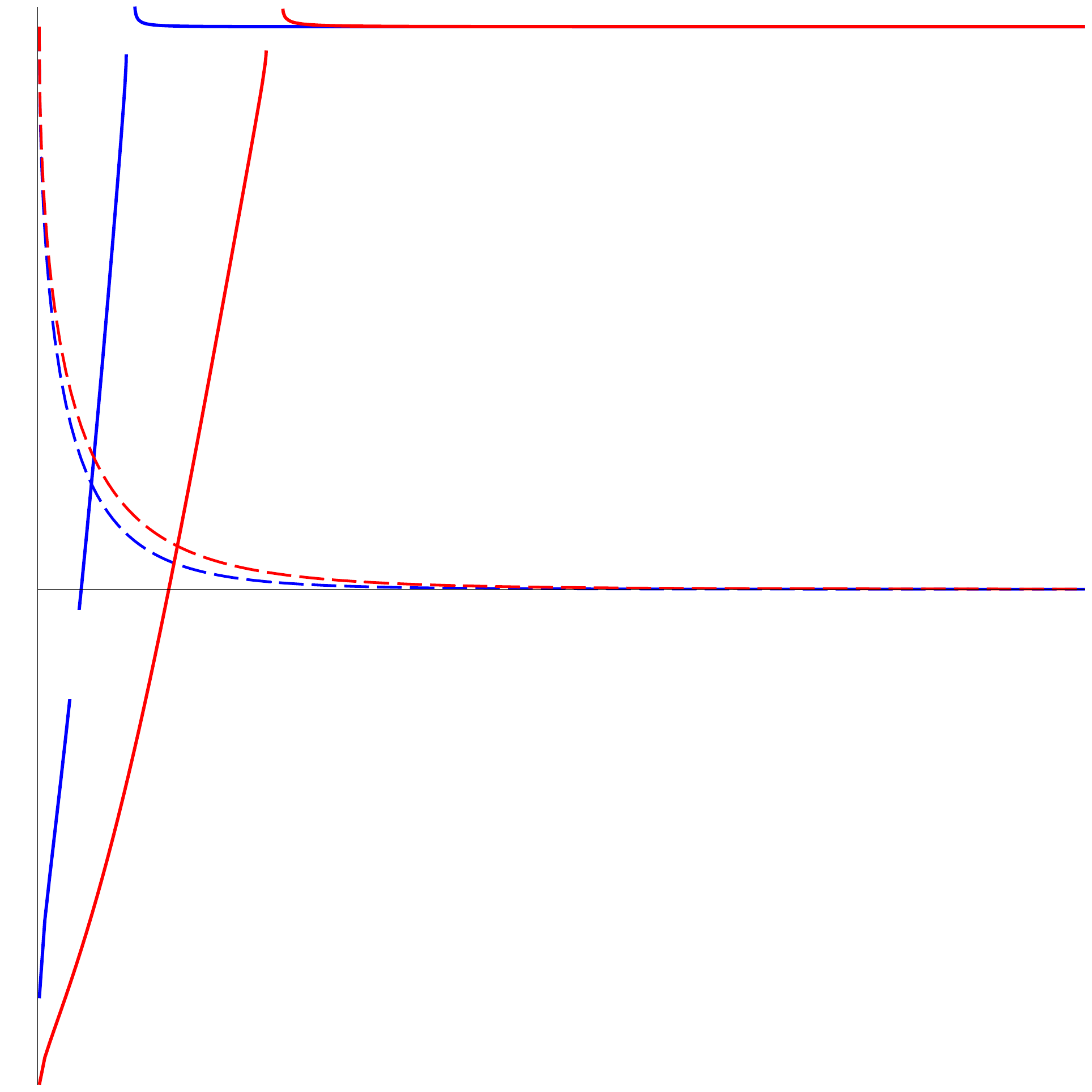}
  \begin{picture}(0,0)(0,0)
\put(-15,86){ $ \mathbf{k}   $}
\put(-135,157){ $ L\Omega_-  $}
\put(-176,66){ $ \hat T  $}
\put(-190,170){ $ 1  $}
\put(-190,77){ $ 0  $}
\end{picture}	
\end{subfigure}%
\begin{subfigure}{.5\textwidth}
  \centering
  \includegraphics[width=.8\linewidth]{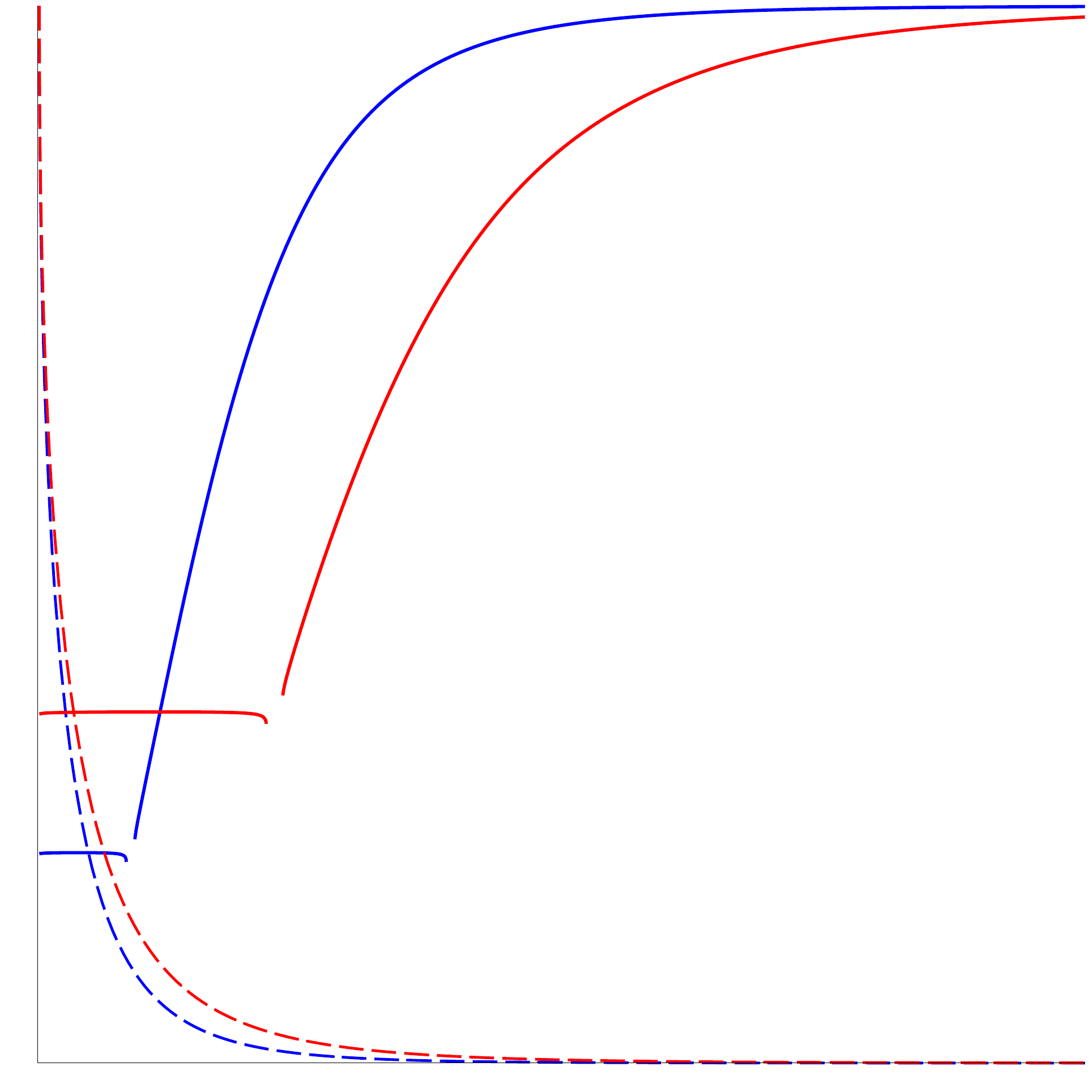}  
  \begin{picture}(0,0)(0,0)
  \put(-106,120){ $ \dfrac{L\Omega_+}{n-2}  $}
   \put(-175,8){ $ \hat T $}
   \put(-15,10){ $ \mathbf k $}
   \put(-193,167){ $ 1 $}
   \put(-193,0){ $ 0 $}
  \end{picture}	
\end{subfigure}
\caption{The angular velocities $\hat \Omega_\pm$ (solid) and relative intrinsic angular momentum $\mathcal S /\mathcal S_{\text{max}}$ (dashed) plotted as a function of $\mathbf{k}$ for the D3 (red) and M5 (blue) thermal giant graviton configurations. The plots are drawn for $\hat T =0.18$ and have been cut off at $\mathbf k = 0.6$ to enhance features.}
\label{omegaplots}
\end{figure}
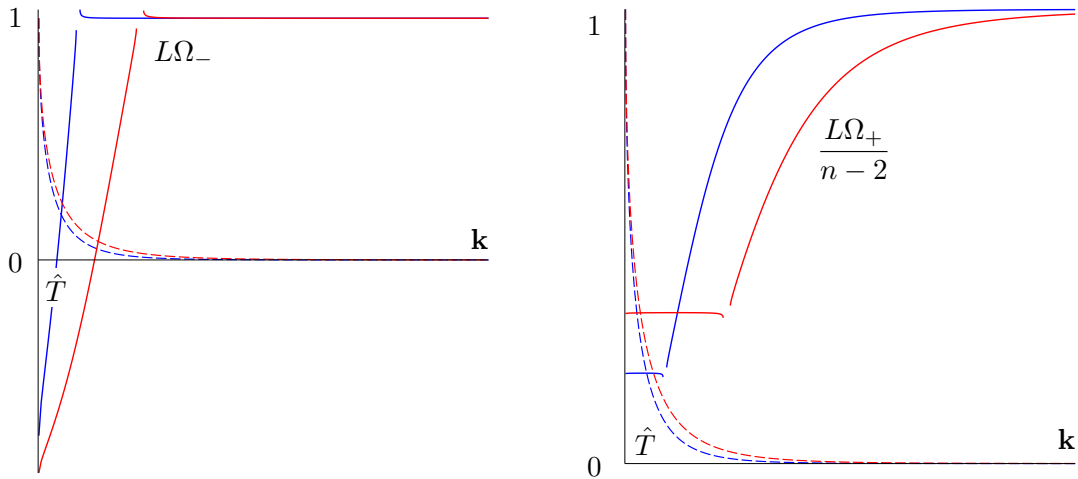

At $r=L$ the Killing vector $\mathbf k$ only depends on $\mathcal W=\hat \omega$. Substituting the expression for $\mathcal W$ in terms of $\mathbf k$ into \eqref{eqq1}, we obtain the solution for $\mathcal V_\pm=\hat \Omega_\pm$ parameterized in terms of $\mathbf k =(1-\hat \omega^2)^{1/2}$ at maximal size. In Fig.~\ref{omegaplots} the angular velocity $\Omega$ is plotted as a function of $\mathbf k$ for both branches for the D3- and M5-brane, respectively. Here we describe the main features of the solutions. 

As can be seen from the plot, there is a small range of values of $\mathbf k$ which admits no solutions to the EOM. Therefore each branch splits up into a  low spin branch and a high spin branch\footnote{This effect can also be deduced by looking at the behavior of the quantity $\mathcal D_\mathcal{W}$ in 
\eqref{eqq2}, see App.~\ref{app:reparameterization}.}. At low spin the angular velocity $\Omega_\pm$ and thermodynamics get small quadratic spin corrections. This is simply because the conserved quantities depend quadratically on the spin parameter $\mathcal W$ except the intrinsic angular momentum which only depends linearly on $\mathcal W$ to lowest order. However, these corrections will be sub-leading to the thermal corrections from the non-zero temperature of the background (see Sec.~\ref{sec:lowtemperature} below).

In the high spin regime the situation is very different and the solution space is dominated by the effects of internal spin. As already pointed out  in Sec.~\ref{sec:solution}, the maximal value for the intrinsic angular momentum is attained as $\mathbf k \to \hat T$. This can also be seen from the plots in Fig.~\ref{omegaplots}. As $\mathbf k$ approaches $\hat T$, we see that the angular velocity $\Omega_-$ crosses zero and becomes negative. In order to examine the solution space near maximal spin we expand around maximal spin $\mathbf k= \hat T(1+\delta^2)$, $\delta \ll 1$. It is straightforward to solve the charge quantization equation \eqref{eq2} to leading order in $\delta$. Notice that for $\mathbf k =\hat T$, we have 
\begin{equation}
\phi\big(\hat T \big)=\frac{m-3}{m-2} \ .
\end{equation}
It is now straightforward to compute the thermodynamics for small $\delta$. For the D3 giant graviton we find 
to leading order in $\hat T$
\begin{equation}
\mathbf E=\frac{2}{\sqrt{3}\thinspace \hat T^2}\left(1-\frac{4\sqrt{2}}{\sqrt{3}}\thinspace \delta+\mathcal{O}\big(\delta^2\big)\right) \ , \quad
\boldsymbol{\mathcal{S}}=\frac{1}{2\sqrt{3} \thinspace \hat T^2}\left(1-\frac{4\sqrt{2}}{\sqrt{3}} \thinspace \delta+\mathcal{O}\big(\delta^2\big) \right)\ .
\end{equation}
Similarly we find for the M5-brane configuration
\begin{equation}
\mathbf E=\frac{1}{\sqrt{2}\thinspace \hat T^2}\left(1-\frac{3\sqrt{3}}{\sqrt{2}} \thinspace \delta+\mathcal{O}\big(\delta^2\big)\right) \ , \quad
\boldsymbol{\mathcal{S}}=\frac{1}{3\sqrt{2} \thinspace \hat T^2}\left(1-\frac{3\sqrt{3}}{\sqrt{2}} \thinspace \delta+\mathcal{O}\big(\delta^2\big)\right) \ .
\end{equation}
Note that to leading order $\hat T \mathbf S$ is of order $\mathcal{O}(\hat T^0)$. To leading order, the free energy is therefore equal to the energy. The above relations can be used to eliminate the small expansion parameter $\delta$ and write the energy in terms of the intrinsic angular momentum in the high spin limit. For the D3 giant graviton, we find (here $\Delta \mathcal S\equiv \mathcal S_{\text{max}}- \mathcal S$)
\begin{equation}
\label{EnD3}
E=\frac{1}{L}\left(\frac{2 \sqrt{2} }{3\cdot 3^{1/4} \pi ^2 }\frac{\sqrt{N^3 N_{\text{D}3}}}{(LT)^2}-4\Delta \mathcal S+\mathcal{O}\big(\Delta \mathcal S^2\big)\right) \ , \ \ \ S_{\text{max}}=\frac{1}{3 \cdot 3^{1/4} \sqrt{2} \pi ^2 }\frac{\sqrt{N^3 N_{\text{D}3}}}{(LT)^2} \ .
\end{equation}
where we have re-introduced the physical units using \eqref{eq:tension} and \eqref{sindelta}. Similarly, we find for the M5-brane
\begin{equation}
\label{EnM5}
E=\frac{1}{L}\left(\frac{9 }{8 \sqrt{2} \pi ^2 }\frac{\left(N^4 N_{\text{M}5}\right)^{1/3}}{(LT)^2}-3\Delta \mathcal S +\mathcal{O}\big(\Delta \mathcal S^2\big)\right) \ , \quad S_{\text{max}}=\frac{3}{8 \sqrt{2} \pi ^2}\frac{\left(N^4 N_{\text{M}5}\right)^{1/3}}{(LT)^2} \ .
\end{equation}
As is clear from the expressions above, the maximally spinning giant graviton configurations are very heavy objects.%
\footnote{We note that the energy in \eqref{EnD3} is proportional to $N^2 (N_{\rm D3}/N)^{1/2} $, while \eqref{EnM5} 
is proportional to $N^{3/2} \lambda_M^{1/6}$ in terms of the 't Hooft like coupling $\lambda_M$ defined below \eqref{validity}.}

\subsection{Low temperature expansion \label{sec:lowtemperature} }
In this section we give an approximate solution to the giant graviton EOMs in terms of the radial coordinate $r$ in a low temperature expansion and without intrinsic spin (this will thus provide the M5- and M2-brane generalizations of Ref.~\cite{Armas:2012bk}). Moreover we briefly examine the low spin and the maximal spin case for a given $r$ in a low temperature expansion, respectively. 

\subsubsection*{The low temperature limit with no intrinsic spin}

In order to work out the low temperature expansion we take $T\to 0$, or equivalently $\phi\to 0$ while keeping $\mathbf k$ finite. First, since $\phi \ll 1$, we can immediately solve the charge conservation equation \eqref{eq2}. Indeed, in this limit the $\phi^{m-2}$ term can be dropped and the solution to \eqref{eq2} is given by 
\begin{equation} \label{lowTphi}
\phi=\mathcal C_m \left(\frac{\hat T}{\mathbf k}\right)^{\gamma_m} \ , 
\end{equation}
where
\begin{equation} \label{lowTCm}
\gamma_m=\frac{2(m-1)}{m-3}  \quad \text{and}  \quad
\mathcal C_m=(m-3)(m-2)^{\tfrac{2-m}{m-3}} \ .
\end{equation}
Notice that for the values of $n$ and $m$ considered in this paper, we have $\gamma_m=\gamma_{D-n}=n-1$. In the limit with 
no intrinsic spin, we therefore find the following solution for $\phi$ 
\begin{equation}
\phi=\phi_0 \thinspace \mathbf{k}^{1-n} \ , \quad \phi_0 \equiv \phi|_{r=L}= f_n \hat{T}^{n-1} \ ,
\end{equation}
where we have defined $f_n\equiv \mathcal C_{D-n}$ and  
\begin{equation}
f_4=\frac{4}{5\cdot 5^{1/4}} \ , \quad f_5=\frac{2}{3\sqrt{3}} \ , \quad f_7=\frac{1}{4}  \ .
\end{equation}
Notice that the limit $\phi \ll 1$ requires that $\mathbf k \gg \hat T$ which is equivalent to $\hat r \gg \hat T$.  We now proceed as in \cite{Armas:2012bk} and expand around the extremal solution \eqref{extremalsolution}. It is straightforward to expand $\mathcal V_\pm$ with $\mathcal W=0$ in terms of $\phi$. One finds\footnote{Note that the expressions and manipulations pertaining to this section only apply to the physical values of $n$ 
and $m$.}
\begin{equation}
\mathcal V_-=\mathbf{k}+\mathcal{O}(\phi^2) \ , \quad \mathcal V_+=(n-2)(1-\phi)\mathbf{k}+\mathcal{O}(\phi^2) \ .
\end{equation}
It is seen that for the physically relevant values of $n$ and $m$, $\mathcal V_-$ gets no first order correction as was also seen in the D3-brane case. Now using
$\mathbf{k}^2=1-\hat \Omega^2 \hat r^2$ and $\mathcal V=\hat r \hat \Omega$, we can solve for $\hat \Omega$
\begin{equation}
\hat \Omega_-\simeq  \hat{\overline{  \Omega}}_-+\mathcal{O}(\phi^2), \quad \hat \Omega_+\simeq\hat{\overline{\Omega}}_+\left(1- \left(\frac{\hat{\overline{\Omega}}_+ \hat r}{n-2}\right)^2\phi\right)+\mathcal{O}(\phi^2) \ .
\end{equation}
where the expressions for the angular velocites $\hat{\overline{  \Omega}}_\pm$ at extremality were recorded in Eq.~\eqref{extremalsolution}. 

It is now possible to compute the on-shell quantities for the lower and upper branch using \eqref{EJ}, \eqref{SS}. For the lower branch, we find 
\begin{equation*}
\mathbf{E}_-\simeq\bar{\mathbf{E}}_-+\frac{n-2}{n-1}\frac{\phi_0}{\hat r^2} \ , \quad  \mathbf{J}\simeq\bar{\mathbf{J}}_-+\frac{n-2}{n-1}\left(\frac{\hat \rho}{\hat r}\right)^2\phi_0 \ , 
\end{equation*}
\begin{equation}\label{lowtempquantitieslower}
\hat{T}\boldsymbol{S}_-\simeq\phi_0 \ , \quad  \mathbf{F}_-\simeq\bar{\mathbf{E}}_--\left(\hat r^2-\frac{n-2}{n-1}\right)\frac{\phi_0}{\hat r^2} \ .
\end{equation}
where $\bar{\mathbf{E}}_-$ and $\bar{\mathbf{J}}_-$ were written down in Eq.~\eqref{EandJextramalminus}. Similarly for the upper branch we find
\begin{equation*}
\mathbf{E}_+\simeq\bar{\mathbf{E}}_++\frac{n-2}{n-1}\!\left(\!\frac{n-2}{\hat{\overline{\Omega}}_+}\!\right)^{\!\!n-2}\!\!\!\left(\!n-1-\frac{n-2}{\hat r^2}\!\right)\phi_0 \!\!~~ , \!\! \ \   \mathbf{J}_+\simeq\bar{\mathbf{J}}_+-\frac{n-2}{n-1}\!\left(\!\frac{n-2}{ \hat{\overline{\Omega}}_+}\!\right)^{\!\!n-1}\!\!\!\left(\frac{\hat \rho}{\hat r}\right)^2\!\!\phi_0 \ ,
\end{equation*}
\begin{equation}\label{lowtempquantitiesupper}
\hat{T}\boldsymbol{S}_+\simeq\left(\frac{n-2}{\hat{\overline{\Omega}}_+}\right)^{\!\!n-2}\!\!\!\!\phi_0 ~~ ,~ ~
\mathbf{F}_+\simeq\bar{\mathbf{E}}_++\frac{n-2}{n-1}\left(\frac{n-2}{ \hat{\overline{\Omega}}_+}\right)^{\!\!n-2}\!\!\!\left(\frac{(n-1)(n-3)}{n-2}-\frac{n-2}{\hat r^2}\right)\phi_0 \ ,
\end{equation}
with $\bar{\mathbf{E}}_+$ and $\bar{\mathbf{J}}_+$ given in \eqref{EandJextramalplus}. If needed, it is easy to reintroduce the dimensions and write the expression in terms of the physical quantities. Simply use that
\begin{equation}
\frac{\phi_0 N N_{\text{M$2$}}}{(LT)^3}=\frac{\sqrt{2}\thinspace 2^5\pi^3}{3^3} N_{\text{M$2$}}^{3/2}~~ , \quad  
 \frac{\phi_0 N N_{\text{D$3$}}}{(LT)^4}=\pi^4 N_{\text{D$2$}}^2~~, \quad 
\frac{\phi_0 N N_{\text{M$5$}}}{(LT)^6}=\frac{2^7 \pi^6 }{3^6} N_{\text{M$5$}}^3~~.
\end{equation}
We now express the free energy $F=E-TS$ on the lower branch in terms of the angular momentum. We find 
\begin{equation}\label{freeenergy}
\begin{split}
F_{\text{M$2$}}&=\frac{J}{L}-\frac{\sqrt{2}\thinspace 2^5\pi^3 }{3^4} N_{\text{M$2$}}^{3/2} L^2T^3 + {\cal{O}} (T^6) \ , \\
F_{\text{D$3$}}&=\frac{J}{L}-\frac{\pi^4  }{4} N_{\text{D$3$}}^{2} L^3T^4 + {\cal{O}} (T^8)  \ , \\
F_{\text{M$5$}}&=\frac{J}{L}-\frac{2^6 \pi^6 }{3^7} N_{\text{M$5$}}^3 L^5T^6 +  {\cal{O}} (T^{12})  \ .
\end{split}
\end{equation}
We observe that, to leading order, the difference $F-J/L$ is proportional to the free energy of the field theories
living on the giant graviton branes  \cite{Klebanov:1996un}. In this connection, we note that it is non-trivial that the 
$J$-dependence  has cancelled out in this difference. 
It is straightforward to write down similar expressions for the upper branch, however, the
resulting expressions involve complicated functions of the angular momentum multiplying
the thermal corrections, so we omit them here. 

Finally, we compute the ratio $J/E$ for the lower branch. We find 
\begin{equation}
\begin{split}
\frac{J_{\text{M$2$}}}{E_{\text{M$2$}}}&=L-\frac{\sqrt{2}\thinspace 2^6\pi^3 L}{3^4 J}  N_{\text{M$2$}}^{3/2} (LT)^3 \ , \\
\frac{J_{\text{D$3$}}}{E_{\text{D$3$}}}&=L-\frac{ 3 \pi^4 L }{4J}  N_{\text{D$3$}}^{2} (LT)^4 \ , \\
\frac{J_{\text{M$5$}}}{E_{\text{M$5$}}}&=L-\frac{5 \cdot 2^6 \pi^6 L }{3^7 J}  N_{\text{M$5$}}^3 (LT)^6 \ .
\end{split}
\end{equation}
The first term is recognized as the usual Kaluza-Klein contribution while the second
term is due to thermal effects coming from the thermal excitations of the $(n-1)$-dimensional field theories living on the giant graviton world volume.

\subsubsection*{The low temperature limit with low intrinsic spin}

Since $\boldsymbol{\mathcal{S}}\sim \phi$ (cf. Eq.~\eqref{SS}), for a given temperature $\hat T$, the scale set for $\boldsymbol{\mathcal{S}}$ is given by $\phi_0$. Let us therefore define 
$\boldsymbol{\mathcal{S}}=s \thinspace \phi_0$. In this way the low spin regime is where $s\ll 1$. In this regime we have $\mathcal W \sim s \ll 1$ and $\mathbf k \simeq |k_{\text{w.v.}}|$. If we further take the low temperature limit, we find to leading order
\begin{equation}
\hat \omega_-=s \ , \quad \hat \omega_+=\left(\frac{\hat{\overline{\Omega}}_+}{n-2} \right)^n s \ .
\end{equation}
In the low temperature regime, the effects of internal spin are first visible to order $\mathcal{O}(\phi_0^2)$. The expression for the conserved quantities \eqref{lowtempquantitieslower} and \eqref{lowtempquantitiesupper} are therefore not changed to leading order.

\subsubsection*{Low temperature and maximal spin case} 

For a given $\hat T \ll 1$, maximal spin is attained for $\mathbf k = \hat T$. Indeed, the lowest possible value for $\mathbf k$ is $\hat T$ (cf. the discussion in Sec.~\ref{sec:solution}). In the low temperature limit, the middle term in the extrinsic equation \eqref{eq2} dominates and therefore $\mathcal V \simeq \mathcal W$. We therefore conclude
\begin{equation}
\hat \omega \simeq \pm \hat \Omega_\pm=1+\mathcal{O}(\hat T^2) \ .
\end{equation}
In the high spin limit we therefore see that the upper and lower branch are on completely the same footing. The upper branch is rotating in the positive direction while the lower branch rotates in the negative direction around the $S^1$. As the intrinsic spin is decreased, the two angular velocities increase so that $\hat\Omega_+$ goes from $1$ to $\hat{\overline{\Omega}}_+$ (+ thermal corrections) and $\hat{\Omega}_-$ goes from $-1$ to $1$ (+ thermal corrections). This behavior can also be seen on the plot in Fig. \ref{omegaplots}. It is easy to work out the maximal spin thermodynamics for any $r\gg \hat T$. One finds the same results as in Sec.~\ref{sec:maxspin} scaled with suitable powers of $\hat r$.

\subsection{Spinning black hole configuration}
Very much as in flat backgrounds, the extrinsic equation allows for stationary $\Omega=0$ odd-sphere
solutions \cite{Emparan:2011hg} (i.e. configurations with only intrinsic spin and $(m,n)=\{(5,5),(4,7)\}$). In order to make connection with Ref.~\cite{Emparan:2011hg} and related works, instead of working in the usual ensemble where we keep $ T$, $r$ and $N_{(n-2)}$ fixed and determine the one parameter space of solutions parameterized by internal spin $\mathcal S$, in this section we keep the size of the giant graviton $r$, the temperature $T$ and the global dipole potential\footnote{Notice that the expression only holds for $\Omega=0$, see \cite{Emparan:2011hg}.} 
\begin{equation}\label{dipolepot}
\Phi_{(n-2)}=\Omega_{(n-2)}r^{n-2}\tanh \alpha \ ,
\end{equation}
fixed. This amounts to simply taking 
\begin{equation}
\alpha_\Phi=\arctan\left(\frac{\Phi_{(n-2)}r^{2-n}}{\Omega_{(n-2)}}\right) \ .
\end{equation}
As we now go along the one parameter family of solutions parameterized by the internal spin $\mathcal S$ at fixed $r$ and $T$, the dipole potential $\Phi_{(n-2)}$ will be constant but the charge $Q_{(n-2)}=T_{(n-2)}N_{(n-2)}$ will vary. For $\Omega=0$, the extrinsic equation \eqref{exteq} takes the simple form
\begin{equation}\label{exteqomegazero}
(n-2)\left(1-\omega_r^2r^2\right)=-\mathcal R_1(\alpha_\Phi)\omega_r^2r^2 \ ,
\end{equation}
with the solution 
\begin{equation}\label{solOmegaequalzero}
\omega_r=\frac{1}{r}\sqrt{\frac{n-2}{n-2-\mathcal R_1(\alpha_\Phi)}} \ , 
\end{equation}
for the internal angular velocity. 

The balancing condition \eqref{solOmegaequalzero} is the same as the one obtained for flat backgrounds \cite{Emparan:2011hg}. This was expected since the coupling to the background $n$-form flux is proportional to $\Omega$ combined with the fact that the extrinsic equation of motion is a local equation. We emphasize that the solution \eqref{solOmegaequalzero} represent a stationary \emph{bona fide} three-parameter\footnote{Described by parameters $(r,r_0,\alpha)$ or through a set of transformations (captured by Eqs.~\eqref{thermoquantities},\eqref{dipolepot}) the physical parameters $(r,T,\Phi_{(n-2)})$.} black hole solution on $\ads_m\times S^n$. Using the formulas \eqref{conservedQ} 
(by substituting $\mathbf k =1-\omega_R^2R^2$ with $\alpha$ fixed), it is possible to obtain the expressions for the black hole mass and thermodynamics in a straightforward manner. However, note that although the balancing condition \eqref{solOmegaequalzero} is equivalent to the balancing equation for odd-sphere solutions in flat backgrounds, the thermodynamics is not the same due to the non-trivial (global) background geometry. In particular the curvature of the $S^n$ will introduce a tension term in the Smarr relation \cite{Armas:2012bk}. Also note that the angular momentum $J$ of these configurations is not vanishing (as it would trivially be in flat backgrounds) due to the presence of the background flux.

If we want to determine the stationary $\Omega=0$ solutions for a given charge $Q_{(n-2)}$ (i.e. switch back to the canonical ensemble), in addition to Eq.~\eqref{solOmegaequalzero} we must also impose \eqref{conservedQ}. This gives an implicit equation for $\omega_r$ which is neither captured by the high spin regime nor the usual low temperature regime. However, it is easy to see that a solution exists by continuity (which can also be seen on the plot in Fig.~\ref{omegaplots}) and obtaining the solution is straightforward numerically. 


\section{Null-wave giant graviton \label{sec:nullwave}}
In this section we examine a specific solution of Eq.~\eqref{eqq1}, consisting of a zero temperature excitation of the usual extremal giant graviton obtained by taking a particular limit for which the fluid velocity becomes light-like. Motivated by this configuration we then write down an action for null-wave branes and show that the result obtained from varying this action and approaching zero temperature in a non-trivial way leads to the same solution. Finally, as an application of this action we obtain the `dual' version of this configuration expanded into $\ads_{m}$.


\subsection{Extremal giant graviton solution with null-wave} \label{sec:extremalnull}

 Here we show that the thermal giant graviton solution obtained in Secs.~\ref{sec:setup} and \ref{sec:thermal} admits a zero-temperature limit which can be regarded as a null-wave excitation of the extremal giant graviton presented in Sec.~\ref{sec:extremal}. This null-wave limit consists in approaching extremality by sending $\phi \to 0$ such that 
\begin{equation} \label{nulllimit}
\frac{\phi}{\mathbf k}=\mathcal P\thinspace\mathbf k~~,~~ \mathbf k  \to 0 \quad \text{($\mathcal P$ fixed)}~~,
\end{equation}
while keeping the charge $Q_{(n-2)}$ constant. This zero-temperature limit is consistent with \eqref{eq2}. Moreover, in this particular limit, the equation of motion \eqref{exteq} simplifies to
\begin{equation}\label{eqnull}
(n-2)\mathcal{W}^2+\mathcal{V}^2-\mathcal{W}^2\mathcal{P}\left(\mathcal{W}^2-\mathcal{V}^2\right)-(n-1) \mathcal{V} \mathcal{W}=0~~.
\end{equation}
The solution to \eqref{eqnull} can also be obtained by taking the appropriate limit \eqref{nulllimit} in the general solution \eqref{eqq1} and reads
\beq \label{nullsol}
\mathcal{V}_{\pm}=\frac{1}{2}\frac{n-1 \pm |n-3-2\mathcal{P}\mathcal{W}^2|}{1+\mathcal{P}\mathcal{W}^2}\mathcal{W}~.
\eeq
As in the extremal case of Sec.~\ref{sec:extremal}~, this results in two branches of solutions
\beq \label{branches}
\hat{\Omega}_{-}=\hat{\omega}\quad,\quad\hat{\Omega}_{+}=\hat{\omega}\left(\frac{n-1}{1+\mathcal{P}\mathcal{W}^2}-1\right)~.
\eeq
The off-shell thermodynamic properties associated with these configurations are obtained from \eqref{conservedQ} together with \eqref{nulllimit} and take the following form:
\beq \label{tfull}
\textbf{E}=\frac{1}{\hat{\omega}}\left(1+\mathcal{P}\hat{\omega}^2\hat{r}^2\right)\hat{r}^{n-3}\quad,
\quad  \hat{T}\textbf{S}=0~~,
\eeq
\beq \label{tfull1}
\textbf{J}=\textbf{E}\hat{\rho}\sqrt{1-\hat{\omega}^2\hat{r}^2}+\hat{r}^{n-1} \quad,\quad
\boldsymbol{\mathcal{S}}_{i}=\frac{2}{n-1}\mathcal{P}\hat{\omega}^2\hat{r}^{n+1} \spa i = 1, \ldots , (n-1)/2 ~~.
\eeq
Contrary to the usual $\frac{1}{2}$-BPS case presented in Sec.~\ref{sec:extremal}, we see that the null-wave giant graviton caries spin along the 
Cartan directions  of the world volume, which vanishes when the momentum density $\mathcal{P}$ vanishes. The null-wave excitation of the extremal giant graviton excites $(n-1)/2$ new extra quantum numbers of equal magnitude. We will now analyze the thermodynamic properties and stability of both branches \eqref{branches} and compare the results with the extremal giant graviton.


\subsubsection*{Lower branch}

 For the branch of solutions $\hat{\Omega}_-$, the requirement that $\textbf{k}=0$ implies that $\hat{\Omega}_-=\hat{\omega}=1$. In fact, this means that not only the center of mass is moving at the speed of light but also all points in the expanded brane. This was not possible for the extremal graviton solution of Sec.~\ref{sec:extremal} as there all brane points are required to move along a timelike Killing vector field. In this case, using Eqs.~\eqref{tfull}-\eqref{tfull1}, the on-shell thermodynamic quantities take the form
\beq \label{thermoquantitiesb}
\textbf{E}=\textbf{E}_{-}+\frac{\boldsymbol{\mathcal{S}}}{\hat{r}^2}\quad,\quad \textbf{J}=\textbf{J}_{-}+\hat{\rho}^2\frac{\boldsymbol{\mathcal{S}}}{\hat{r}^2}~,
\eeq
where $\textbf{E}_{-}$ and $\textbf{J}_{-}$ denote the energy and angular momentum of the lower branch extremal giant graviton given in Sec.~\ref{sec:extremal} and $\boldsymbol{\mathcal{S}}$ denotes the sum of all the spins, i.e., 
\beq
\boldsymbol{\mathcal{S}}=\sum_{i=1}^{(n-1)/2}\boldsymbol{\mathcal{S}}_i=\mathcal{P}{\hat r}^{n+1}~~. 
\eeq
These relations are of particular interest as they indeed show that this configuration can be seen as a zero-temperature excitation of the lower branch of the extremal giant graviton. Furthermore, from \eqref{thermoquantitiesb} we obtain the relation:
\beq \label{EJb}
\textbf{E}=\textbf{J}+\boldsymbol{\mathcal{S}}~.
\eeq
This relation is also interesting in its own right as it shows, in the case of $\ads_{5}\times S^{5}$, that we are dealing with a configuration with a $\frac{1}{8}$-BPS spectrum since it satisfies the expected BPS bound $\textbf{E}=\textbf{J}+\boldsymbol{\mathcal{S}}_1+\boldsymbol{\mathcal{S}}_2$. Similarly, in the case of $\ads_{4}\times S^{7}$ it corresponds to a configuration with a $\frac{1}{16}$-BPS spectrum since $\textbf{E}=\textbf{J}+\boldsymbol{\mathcal{S}}_1+\boldsymbol{\mathcal{S}}_2+\boldsymbol{\mathcal{S}}_3$. If the giant graviton has maximal size, $\hat r=1$, the BPS relation \eqref{EJb} simplifies to $\textbf{E}=\textbf{J}+\mathcal{P}$.


\subsubsection*{Upper branch and comparison between branches} 

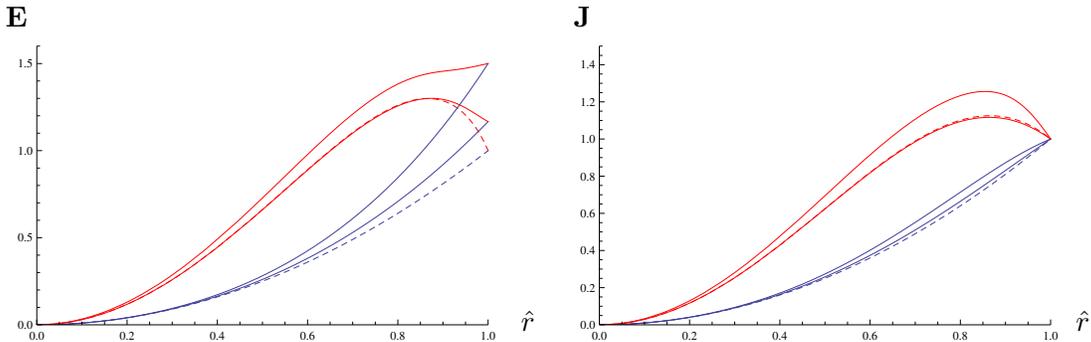
\begin{figure}[!ht]
\centerline{\includegraphics[scale=0.5]{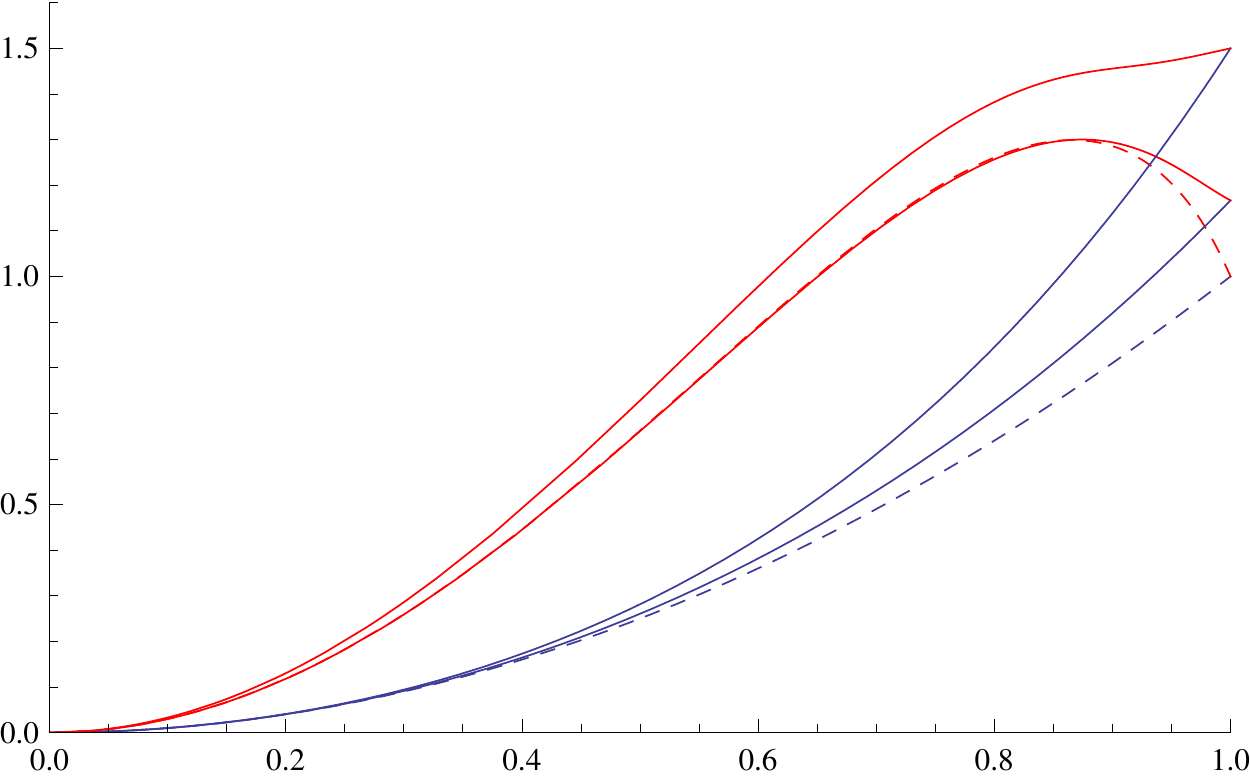} \hskip 1cm
\includegraphics[scale=0.5]{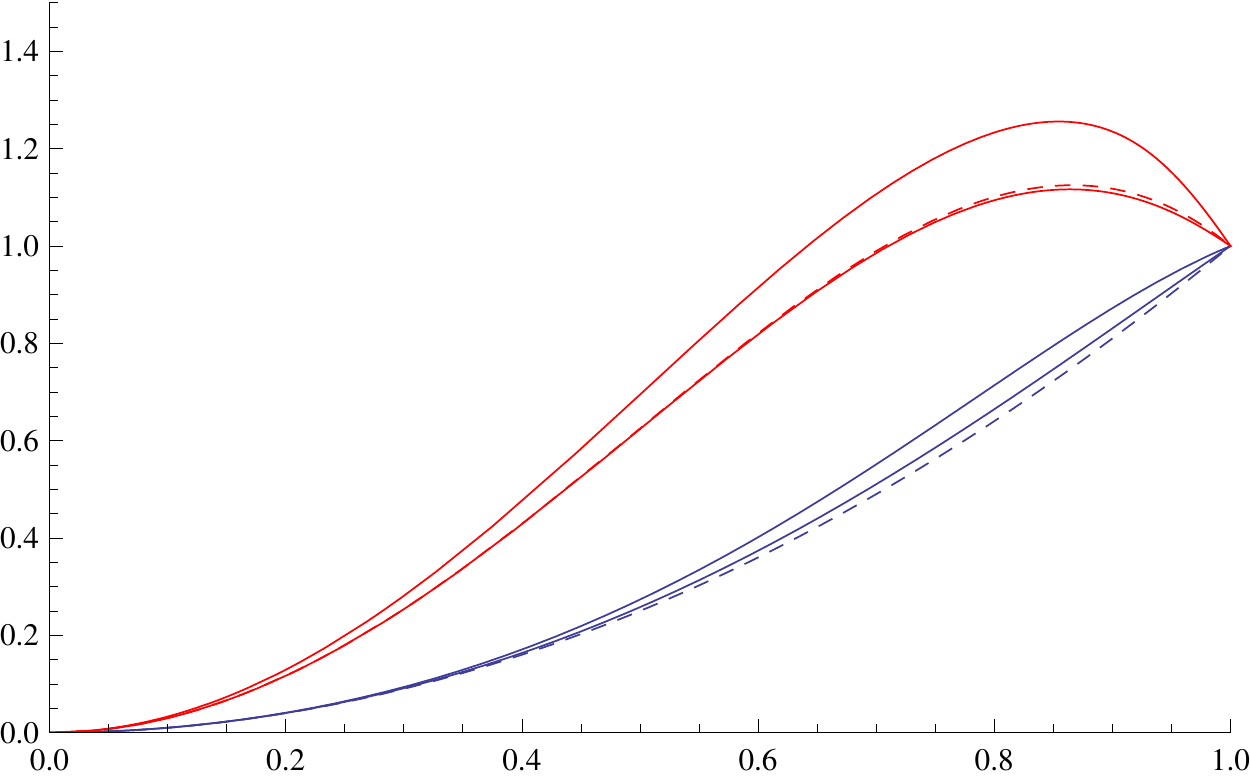} }
\begin{picture}(0,0)(0,0)
\put(210,20){ $\hat r $}
\put(15,135){ $ \mathbf{E}   $}
\put(420,20){ $\hat r $}
\put(230,135){ $ \mathbf{J}  $}
\end{picture}		
\vskip -.5cm
	\caption{$\mathbf{E}$  versus  $\hat r $ (left plot) and $\mathbf{J}$ versus $\hat r$ (right plot) for $\mathcal{P}=0,~\frac{1}{6},~\frac{1}{2}$ and $n=5$~. The dashed lines represent the plots for the extremal giant graviton with $\mathcal{P}=0$ while the uppermost curve represents the case $\mathcal{P}=\frac{1}{2}$~.} 
	\label{fig:Evsr}
\end{figure}

For the upper branch solution $\Omega_+$, one can also solve the constraint $\textbf{k}=0$~, however the resulting expression for $\omega$ is too cumbersome to be presented here. Nevertheless, in the limit in which $\mathcal{P}$ vanishes the constraint $\textbf{k}=0$ yields the value of $\hat{\omega}$
\beq \label{omegaup}
\hat{\bar{\omega}}=\frac{1}{\sqrt{(n-2)^2-(n-3)(n-1)\hat{r}^2}}~,
\eeq
which when inserted into \eqref{tfull} gives rise to the thermodynamic properties of the upper branch of the extremal giant graviton as given in Sec.~\ref{sec:extremal}. The upper branch solution in \eqref{branches} has generically a non-BPS spectrum for all values of $\mathcal{P}$ except when the giant graviton acquires maximal size. This is clear when looking at Fig.~\ref{fig:Evsr}, since for all values of $\mathcal{P}$ the two branches meet at $\hat{r}=1$ and therefore the charges $\mathbf{E}$ and $\mathbf{J}$ are equal at maximum size. These plots are obtained by solving the constraint $\textbf k=0$ for the upper branch and obtaining $\hat{r}\left(\mathcal{P}\right)$. The bound on $\hat{r}$, i.e., $0<\hat{r}\le1$ implies the bound $\frac{1}{3}\le\hat\omega\le1$ on $\hat\omega$. These bounds in turn imply that at maximality the total spin $\boldsymbol{\mathcal{S}}$ is equal for both branches. In contrast with the thermal spinning case analyzed in Sec.~\ref{sec:thermal} the spin of these null-wave giant graviton configurations is not bounded from above and from Eqs.~\eqref{thermoquantitiesb} neither is the energy nor the orbital angular momentum. Fig.~\ref{fig:Evsr} also shows that indeed, the configuration characterized by \eqref{branches} is a deformation of the extremal giant graviton (dashed line).
\vskip .7cm
%


\subsubsection*{Stability} 

To study the stability of the solution branches \eqref{branches} we employ the method used in \cite{Armas:2012bk} which consists in considering the thermodynamic ensemble parametrized by the size $r$, the conserved orbital angular momentum $\mathbf{J}$, the conserved spins $\boldsymbol{\mathcal{S}}_i$ and the conserved total charge $Q_{(n-2)}$, and looking for the configurations that minimize the energy $\textbf{E}$. A small off-shell perturbation along $r$ of the angular velocity $\omega$ and the momentum density $\mathcal{P}$, with $\mathbf{J},~\boldsymbol{\mathcal{S}}_i$ and $Q_{(n-2)}$ held fixed, allows us to determine the second derivative of $\textbf{E}$ with respect to $r$. For the lower branch this takes the simple form:
\beq
\textbf{E}_{(2)}^{-}=\frac{1}{2}\frac{(n-3-2\mathcal{P}\hat{r}^2)^2}{(1-\hat{r}^2)(1+\mathcal{P}\hat{r}^2)}\hat{r}^{n-3}~.
\eeq
In the case $\mathcal{P}=0$ and $n=5$ we recover the second variation of the energy for the lower branch extremal giant graviton \cite{Armas:2012bk}. If we restrict to the cases $\mathcal{P}>0$, as otherwise the energy would be negative (see Eq.~\eqref{tfull}), we always have that $\textbf{E}_{(2)}^{-}>0$. This means that the lower branch of the null-wave giant graviton is always stable as expected for BPS configurations. In the case of the upper branch, expanding around the extremal value \eqref{omegaup} for $n=5$ one obtains
\beq
\textbf{E}_{(2)}^{+}=\frac{\hat{r}^2}{2\hat{\rho}^2}\left(4\bar{\Omega}_{+}(r)\left(\frac{4}{3}\hat{r}^2-1\right)+\frac{(9-28\hat{r}^2+16\hat{r}^4)}{\hat{\rho}^2}\tilde\omega\right)~~,
\eeq
where we have introduced the expansion parameter $\tilde\omega=\omega-\hat{\bar{\omega}}$. In the case for which $\tilde\omega=0$~,one recovers the result for the extremal giant graviton, namely, that the upper branch is only stable if $\hat{r}>\hat{r}_*$ where $\hat{r}_*=\sqrt{3}/2$ \cite{Armas:2012bk}. As the spin parameter $\hat\omega$ is introduced the value of $\hat r_*$ can be determined numerically and increases with increasing $\hat\omega$. The range of stability of the upper branch is decreased for increasing spin. This feature is also seen in the case of the giant graviton constructed from wrapping an M5-brane around the $S^{7}$ of $\ads_{4}\times S^{7}$.
\vskip .7cm
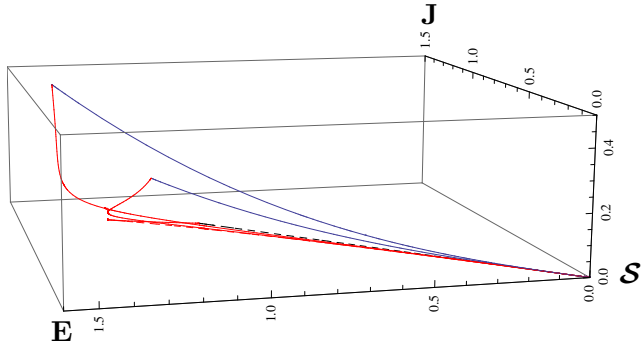
\begin{figure}[!ht]
\centerline{\includegraphics[scale=0.8, angle=-270]{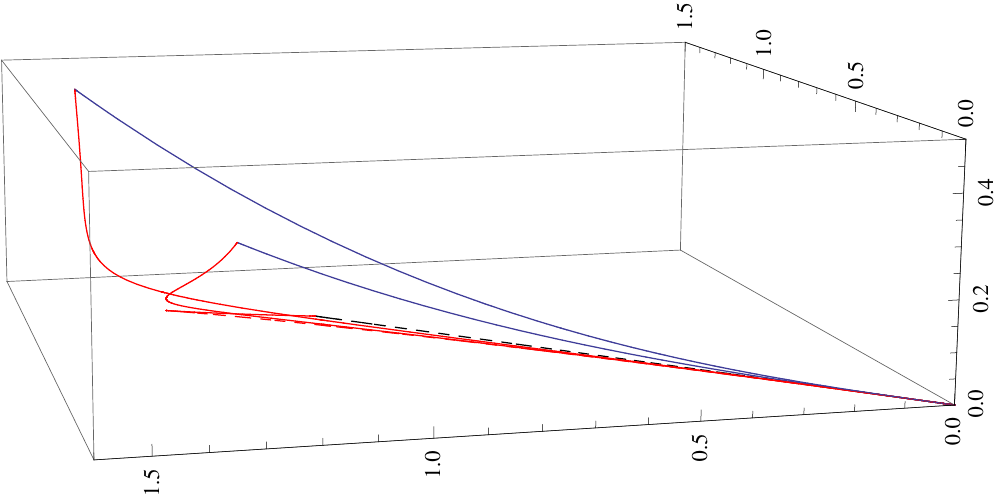}}
\begin{picture}(0,0)(0,0)
\put(335,35){ $\boldsymbol{\mathcal{S}}$}
\put(120,13){ $ \mathbf{E}   $}
\put(260, 133){ $ \mathbf{J}   $}
\end{picture}		
\vskip -.5cm
	\caption{$\mathbf{E}$ versus $\boldsymbol{\mathbf{S}}$ and $\mathbf{J}$ for the values $\mathcal{P}=0,~\frac{1}{6},~\frac{1}{2}$ and $n=5$~. The dashed line represents $\mathcal{P}=0$ while the uppermost curve represents $\mathcal{P}=\frac{1}{2}$~.
	}
	\label{fig:EvsJS}
\end{figure}
These results could have been anticipated by looking at Fig.~\ref{fig:EvsJS}. For each value of $\boldsymbol{\mathbf{S}}$ the surface intersecting the curve for fixed $\mathcal{P}$ selects two different values of $\mathbf{J}$. The value of $\mathbf{J}$ corresponding to the lowest energy $\mathbf{E}$ is the one corresponding to the lower branch in \eqref{branches}. However, if $\mathbf{J}$ is increased beyond the BPS bound, the lower branch ceases to exist and the stable configurations lie within the stable region of the upper branch solution $\hat r_*\le\hat r\le1$. This is a very similar picture to the stability properties of the $\mathcal{P}=0$ extremal giant graviton \cite{Armas:2012bk}.

 
 \subsection{Action for null-wave branes \label{sec:null action}}
In this section we obtain an action for null-wave branes by taking an appropriate limit of the action \eqref{action}. We begin by stressing that the extremal limit of \eqref{action} that yields  the DBI action multiplied by a factor of $N_{(n-2)}$ is obtained by sending $r_{0}\to0$ and $\alpha\to\infty$ such that the total charge $Q_{(n-2)}$ is held constant. Equivalently, using the parameter $\phi$ introduced in Sec.~\ref{sec:setup}, the same limit is obtained by sending $\phi\to0$. However, we are interested in near-extremal situations for which $\phi$ is taken to be small but non-zero. In these cases, the fluid pressure approaches $P\to-Q_{(n-2)}(1-\phi/(n-1))$. Using now the low temperature expansion obtained in Eqs~.\eqref{lowTphi}-\eqref{lowTCm} as $\phi\to0$, the action \eqref{action} reduces to\footnote{We have written the action \eqref{neaction} adapted to the background space-time and configurations studied here but we stress that this action is easily generalized for any other background and for the large class of branes studied in \cite{Emparan:2011hg}.}
\beq\label{neaction}
I=-Q_{(n-2)}\int_{\mathcal{W}_{n-1}}\!\!\!d^{n-1}\sigma\sqrt{-\gamma}\left(1-\frac{f_n}{n-1}\Big(\frac{\hat T}{\textbf{k}}\Big)^{n-1}\right)+\int_{\mathcal{W}_{n-1}}\mathbb{P}[A_{[n-1]}]~~.
\eeq
In the case for which the temperature is taken to zero, the action \eqref{neaction} reduces to $N_{(n-2)}$ times the DBI action plus the Wess-Zumino contribution. When the temperature is non-zero, it accounts for near-extremal excitations of ground state configurations. Noting that by definition $\textbf{k}=|-\gamma_{ab}\textbf{k}^{a}\textbf{k}^{b}|^{\frac{1}{2}}$, the world volume stress tensor of the excited state can be obtained from \eqref{neaction} in the usual way \cite{Armas:2012jg} and takes the form
\beq \label{stne}
T^{ab}=Q_{(n-2)}f_n\Big(\frac{\hat T}{\textbf{k}}\Big)^{n-1}\left(u^{a}u^{b}+\frac{1}{n-1}\gamma^{ab}\right)-Q_{(n-2)}\gamma^{ab}~~.
\eeq
From the form of the world volume stress tensor it is clear that as $\hat T\to0$ we obtain the known result for Dirac branes at zero temperature. 

The expression \eqref{stne} suggests the existence of a scaling limit as $\hat T\to0$ different from the usual extremal limit \cite{Emparan:2011hg}. This is obtained by sending $\hat T\to0$ while the fluid velocity approaches the speed of light $\textbf{k}\to0$ such that $\sqrt{f_n}(\hat T/\textbf{k})^{\frac{n-1}{2}}u^{a}\to\sqrt{\mathcal{P}}l^{a}$ for constant $\mathcal{P}$. In this case, the world volume stress tensor of the excitation is given by
\beq \label{nullstress}
T^{ab}=\mathcal{K}\thinspace l^{a}l^{b}-Q_{(n-2)}\gamma^{ab}~~,
\eeq
where we have introduced the momentum density $\mathcal{K}$ via the relation $\mathcal{K}=Q_{(n-2)}\mathcal{P}$ and also the null-vector $l^{a}$ satisfying $l^{a}l_{a}=0$\footnote{The world volume stress tensor \eqref{nullstress} can also be obtained by taking the equivalent limit $r_0\to0$ and $\textbf{k}\to0$ such that $(\Omega_{(n+1)}nr_0^n)^{\frac{1}{2}}\textbf{k}^{a}=(16\pi G\mathcal{K})^{\frac{1}{2}}\textbf{k}\thinspace l^{a}$ \cite{Emparan:2011hg}.}. The world volume stress tensor \eqref{nullstress} is that of a null-wave: a zero-temperature excitation of the Dirac brane world volume stress tensor carrying a conserved momentum current along a null-vector $l^{a}$. When the momentum density $\mathcal{K}$ vanishes, one obtains the result for Dirac branes. For the case of non-zero $\mathcal{K}$, the near-extremal action \eqref{neaction} can be exchanged by a simpler one for which the variational principle holds the momentum density $\mathcal{K}$ constant instead of the temperature $T$\footnote{Note that the variational principle also holds the charge $Q_{(n-2)}$ constant since $D_{a}Q_{(n-2)}=0$ and hence $\mathcal{P}$ is also held constant. Further, in order to write \eqref{nullaction} we have used the fact that the variation of $\delta\phi$ is given by $\delta\phi=-(1/\gamma_m)\phi\delta\log\textbf{k}$~. Furthermore, the action \eqref{nullaction} is general for all $p$-branes studied in \cite{Emparan:2011hg} and for any background space-time if one simply replaces $n$ by $p+2$~.},
\beq\label{nullaction}
I=-Q_{(n-2)}\int_{\mathcal{W}_{n-1}}\!\!\!d^{n-1}\sigma\sqrt{-\gamma}\left(1+\frac{1}{2} \mathcal{P}\thinspace \textbf{k}^2\right)+\int_{\mathcal{W}_{n-1}}\mathbb{P}[A_{[n-1]}]~~.
\eeq
The world volume stress tensor \eqref{nullstress} then follows from \eqref{nullaction} by first obtaining it for general $\textbf{k}$ and afterwards taking the limit $\textbf{k}\to0$. The equations of motion that follow by varying \eqref{nullaction} take the form \cite{Armas:2012bk}
\beq \label{bfeom}
D_{a}T^{ab}=0~~~,~~~T^{ab}{K_{ab}}^{\mu}=\frac{1}{(n-1)!}\perp{^{\mu}}_{\nu}F^{\nu\rho_1...\rho_{n-1}}J_{\rho_1...\rho_{n-1}}~~,
\eeq
where ${K_{ab}}^{\mu}$ is the extrinsic curvature of the embedding surface, $\perp{^{\mu}}_{\nu}$ projects orthogonally to the world volume directions and $F_{[n]}=dA_{[n-1]}$ is the background field strength\footnote{Here we have assumed that the force term on the r.h.s. of the second equation in Eq.~\eqref{bfeom} does not work on the world volume. The reader should see Ref.~\cite{Armas:2012bk} for more details on how to compute these quantities.}. Here note that the first equation in \eqref{bfeom} is trivially satisfied as a consequence of stationarity \cite{Armas:2013hsa} and the only non-trivial dynamics are encoded in the second equation of \eqref{bfeom}. When introducing \eqref{nullstress} into \eqref{bfeom} leads to Eq.~\eqref{eqnull} for the particular embedding geometry of the giant graviton.


\subsubsection*{Conserved momentum current and spin}
The equations of motion \eqref{bfeom} that arise by varying the action \eqref{nullaction} express conservation of the world volume stress tensor \eqref{nullstress} along world volume directions and balance of mechanical forces along transverse directions to the world volume. However, the first equation in \eqref{bfeom} now splits into two equations
\beq
l^{b}D_{b}l^{a}=0~~,~~D_{a}\left(\mathcal{K}l^{a}\right)=0~~.
\eeq
The first equation above requires the null vector $l^{a}$ to generate geodesics along the world volume while the second equation expresses the conservation of the momentum current. The momentum current can be integrated in order to obtain a conserved momentum charge associated with the near-extremal configuration. However this charge is not independent and is related to the existence of angular momenta along world volume directions (spin) of the configuration. Indeed, for the configurations presented in the previous sections, the spin along the world volume Killing vector field $\chi_{i}$ can be evaluated using the expression
\beq \label{spinformula}
\mathcal{S}_{i}=\mathcal{K}\int_{\mathcal{B}_{n-2}}d^{n-2}\sigma\sqrt{-\gamma}\thinspace l_{a}\chi^{a}_{i}~~,
\eeq
where $\mathcal{B}_{n-2}$ is the spatial part of the world volume. If the momentum density $\mathcal{K}$ vanishes, the configuration carries no spin. Using \eqref{spinformula} results in the value for the spin written in \eqref{tfull}. The energy and angular momentum along transverse directions to the world volume can be evaluated using the formulae given in \cite{Armas:2012bk} together with the world volume stress tensor \eqref{nullstress}.

 
\subsection{Null-wave giant graviton expanded into $\ads_{m}$ \label{sec:nwads}}
Here we obtain the `dual' version of the spinning giant graviton configuration of Sec.~\ref{sec:extremalnull}, namely of
$(m-2)$-branes
 expanded into the $S^{m-2}$ sphere of the $\ads_{m}$ part of the space-time (but still moving on a circle in $S^n$) using the action \eqref{nullaction}. We begin by parameterizing the $\ads_{m}$ metric as
\beq
ds^2_{\ads_{m}}=-R_0^2\thinspace dt^2+R_0^{-2}d\tilde\rho^2+\tilde\rho^2d{\Omega^2_{(m-2)}}~~,~~R_0^2=1+\frac{\tilde\rho^2}{\tilde L^2}~~,
\eeq
 where $\tilde L=2L/(m-3)$ and the metric on the sphere \eqref{spherecoord} is parametrized by the coordinates $\alpha_i$~. The giant graviton is now placed at $\tilde \rho=r$ while the background gauge field with support on the $S^{(m-2)}$ takes the form
 \begin{equation}
A_{[m-1]} =-\frac{r^{n-1}}{\tilde L} dt \wedge d \Omega_{(m-2)}~~.
\end{equation}
The world volume Killing vector field is in this case $|k_{\text{w.v.}}|^2=R_0^2-\Omega^2L^2$~, while the fluid velocity is $\textbf{k}^2=|k_{\text{w.v.}}|^2-\omega^2r^2$. The action \eqref{nullaction} takes the simple form
\beq
\beta I_{\text{E}}=Q_{(n-2)}\Omega_{(m-2)}r^{m-2}\left[|k_{\text{w.v.}}|\left(1+\frac{1}{2} \mathcal{P}\thinspace \textbf{k}^2\right)-\frac{r}{\tilde L}\right]~~.
\eeq
Explicit variation and taking the limit $\textbf{k}\to0$ leads to the equation of motion
\beq
(m-2)\mathcal{W}^2\tilde L^2+r^2+\mathcal{W}^2\mathcal{P}(1-\omega^2\tilde{L}^2)r^2-(m-1)\mathcal{W} \tilde L r=0~~.
\eeq
 This equation admits two branches of solutions as its `dual' version in Sec.~\ref{sec:extremalnull}. However, the upper branch of solutions is less interesting as it is never BPS. This is in fact the same feature observed for the upper branch of the usual $\frac{1}{2}$-BPS giant graviton \cite{Armas:2012bk}. Our focus will be on the lower branch of solutions which takes the simple form of
 \beq \label{nullads}
\hat \Omega_{-}=1~~,~~\hat\omega=1~~,
 \eeq
 where we have rescaled $\Omega$ and $\omega$ such that $\hat{\Omega}=\Omega L$ and $\hat\omega=\omega\tilde L$. 
 
 \subsubsection*{Thermodynamic properties and stability}
 
\begin{figure}[!ht]
\centerline{\includegraphics[scale=0.5]{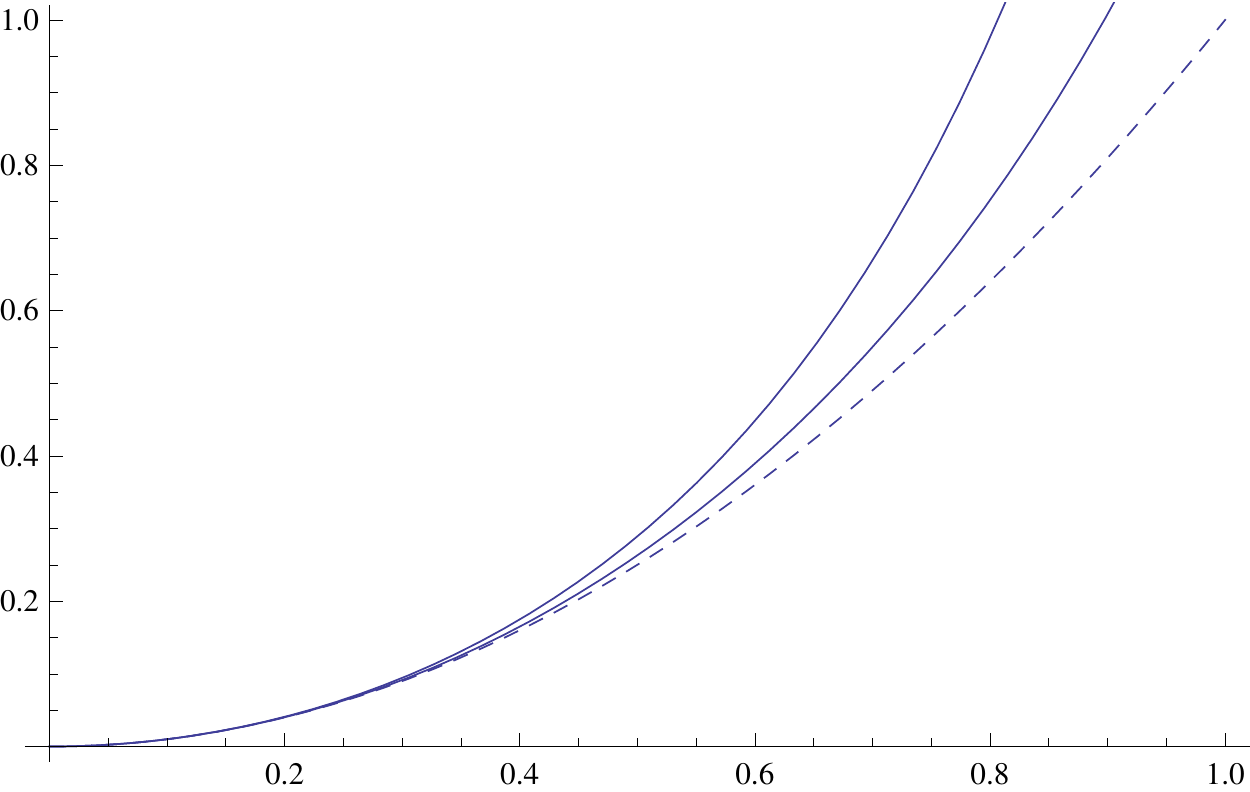} \hskip 1cm
\includegraphics[scale=0.5]{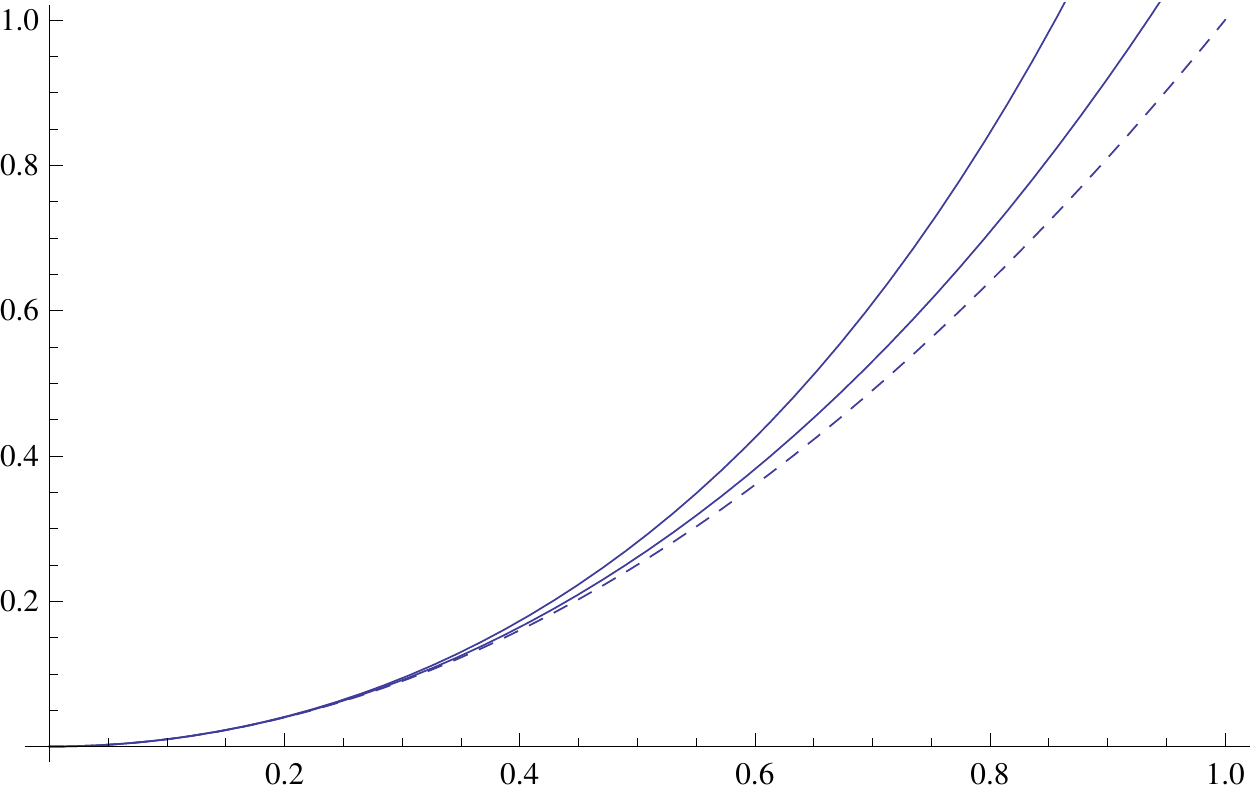} }
\begin{picture}(0,0)(0,0)
\put(210,20){ $\hat r $}
\put(15,135){ $ \mathbf{E}   $}
\put(420,20){ $\hat r $}
\put(230,135){ $ \mathbf{J}  $}
\end{picture}		
\vskip -.5cm
	\caption{$\mathbf{E}$  versus  $\hat r $ (left plot) and $\mathbf{J}$ versus $\hat r$ (right plot) for $\mathcal{P}=0,~\frac{1}{6},~\frac{1}{2}$ and $n=5$ for the lower branch of solutions. The dashed lines represent the plots for the extremal giant graviton with $\mathcal{P}=0$ while the uppermost curves correspond to the case $\mathcal{P}=\frac{1}{2}$~. The plot was restricted to the range $0\le\hat{r}\le1$.} 
	\label{fig:EvsrAds}
\end{figure}
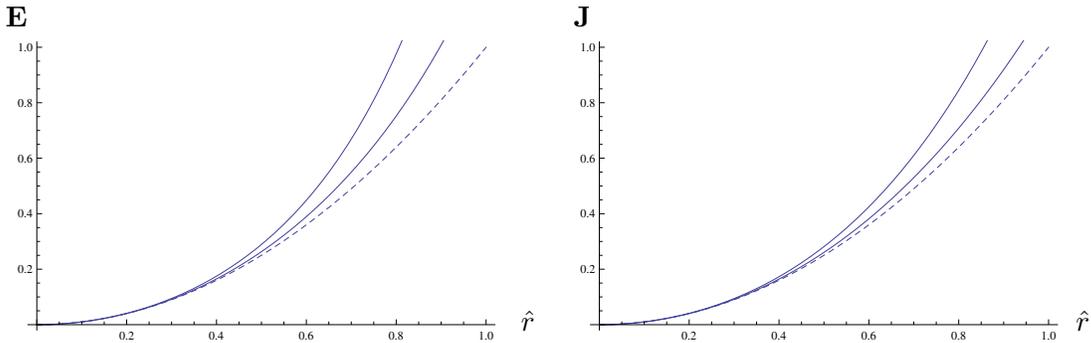

  Using the formulae for thermodynamic quantities given in \cite{Armas:2012bk} and Eq.~\eqref{spinformula} for the spin of the configuration we obtain the following off-shell expressions\footnote{Here we have introduced the ratio $\hat{r}=r/\tilde L$ as well as the rescaled quantities
\beq\nonumber
\textbf{E}=\frac{E}{\Omega_{(m-2)}Q_{(m-2)}\tilde L^{m-2}}~,~\textbf{J}=\frac{J}{\Omega_{(m-2)}Q_{(m-2)}\tilde L^{m-2}L}
~,~ \boldsymbol{\mathcal{S}}=\frac{\mathcal{S}}{\Omega_{(m-2)}Q_{(m-2)}\tilde L^{m-1}}~.
\eeq}
 \beq
 \textbf{E}=\frac{1}{\hat\omega}R_0^2\thinspace\left(1+\mathcal{P}\hat{\omega}^2\hat{r}^2\right)\hat{r}^{m-3}-\hat{r}^{m-1}~~,~~\hat{T}\textbf{S}=0~~,
 \eeq
 \beq
\textbf{J}=\frac{1}{\hat\omega}\sqrt{1+\hat{r}^2(1-\hat{\omega}^2)}\left(1+\mathcal{P}\hat{\omega}^2\hat{r}^2\right)\hat{r}^{m-3}~~,~~ \boldsymbol{\mathcal{S}}_i=\frac{2}{n-1}\mathcal{P}\hat\omega^2\hat{r}^{m+1}~~.
 \eeq
 For the specific solution \eqref{nullads} one can find the relations
 \beq
 \textbf{E}=\textbf{E}_{-}+(1+\hat{r}^2)\frac{\boldsymbol{\mathcal{S}}}{\hat{r}^2}~~,~~\textbf{J}=\textbf{J}_{-}+\frac{\boldsymbol{\mathcal{S}}}{\hat{r}^2}~~,
 \eeq
 implying the BPS bound $\textbf{E}=\textbf{J}+\boldsymbol{\mathcal{S}}$, where $\boldsymbol{\mathcal{S}}$ is the sum over all the $(n-1)/2$ spins. The effect of increasing the spin on the energy and angular momentum can be seen by looking at Fig.~\ref{fig:EvsrAds}.
 \vskip .7cm
As the spin is increased both the energy and angular momentum increase for fixed $\hat{r}$. Fig.~\ref{fig:EvsrAds} depicts the interval $0\le\hat{r}\le1$ but we note that for these configurations in which the giant graviton is expanded into the $\ads_m$ part, the size $r$ is unbounded from above. The stability properties can be analyzed using the method outlined in Sec.~\ref{sec:extremalnull}. In this case we find for the second variation of the energy
on the lower branch
\beq
\textbf{E}_{(2)}^{-}=\frac{1}{2}\frac{(m-3-2\mathcal{P}\hat{r}^2)^2}{(1+\hat{r}^2)(1+\mathcal{P}\hat{r}^2)}\hat{r}^{m-3}~.
\eeq
Therefore we see that these configurations are always stable for any value of $r$ as expected for BPS configurations.

\section{Discussion and outlook \label{sec:outlook}}

In this paper we have constructed and analyzed thermal spinning  giant gravitons in both type II string theory and M-theory.
For extremal giant gravitons, at zero temperature, the world volume stress tensor is Lorentz invariant, so internal spin on the sphere
is a gauge degree of freedom and hence ``invisible''. Heating up the giant graviton breaks the Lorentz invariance, allowing for the introduction of new quantum numbers, namely, the internal spin. The results of \cite{Armas:2012bk} and the present paper, show that by thermalizing giant gravitons 
(in the supergravity regime) we find interesting finite temperature objects in supergravity exhibiting a variety of new qualitative and quantitative effects,
while at the same time we gain access to connections between extremal and null-wave objects. 

We emphasize that  the thermal spinning giant gravitons we have constructed, consisting of the background  together with the 
thermal probe brane placed in it, are \emph{bona fide} solutions of the supergravity equations of motion, to leading order in the blackfold limit. 
This is even true for high  temperatures (i.e. also above the Hawking-Page temperature) as long as $T \leq T_{\rm max}$, provided that we are in the regime of validity in which the black brane can be treated as a probe (see Sec.~\ref{sec:solution}) . However, it would be interesting to see what happens to our solutions
when heated up beyond the Hawking-Page temperature by repeating the analysis for the corresponding $\ads$ black hole backgrounds.

As mentioned above, the thermal spinning giant gravitons of this paper are leading order classical solutions of the relevant supergravity theory. 
However, the blackfold approach provides
 a well-defined scheme in which one can consider higher-order corrections. 
 It would be interesting in this context to study higher-order (elastic) corrections%
 \footnote{Viscuous corrections of (charged) black branes have been considered in 
  \cite{Camps:2010br,Emparan:2013ila}  where in particular Ref.~\cite{Emparan:2013ila} considered D3-branes showing that
  the blackfold effective field theory approach subsumes the constructions encountered in the fluid/gravity correspondence and the black hole membrane paradigm.}
  using the results of  
 \cite{Armas:2011uf,Camps:2012hw,Armas:2012jg,Armas:2013hsa}. 
 For the D3-brane case this would reveal what happens for larger values of the perturbative ratio $N_{\rm D3}/N$, while for the M-brane cases this would involve perturbative effects involving the 't Hooft like coupling $\lambda_M$ (identified in Ref.~\cite{Niarchos:2012cy}, see Sec.~\ref{sec:solution}) with 
 corrections governed by $\lambda_M $ for M5-branes and $1/\lambda_M$ for M2-branes. 
 Furthermore, we recall that as a byproduct of our analysis we have found, in the blackfold limit, new stationary dipole-charged black hole solutions 
 with horizon topology $S^m \times S^{n-2}$ in
$\ads_m \times S^n$ type II/M-theory backgrounds for ${m,n}=(5,5)$ and $(4,7)$. It would be interesting to examine these further,  and perhaps construct the full solution numerically. 
 Another aspect that deserves deeper study is the supersymmetry of the null-wave giant gravitons, which would be first of all important to verify explicitly to leading order, and subsequently examine at higher orders. 

We have considered in this paper the maximally symmetric spinning case with equal angular velocity on each of the Cartan directions. It could be interesting to study the less symmetric case with arbitrary angular velocities (see App.~A of \cite{Emparan:2009vd} where  this was
studied for stationary $S^3$-blackfolds in asymptotically flat space). Another interesting generalization is to construct the spinning 
thermal giant gravitons on products of odd-spheres, in analogy with \cite{Emparan:2009vd,Emparan:2011hg}.  This
would involve M2-branes on $T^2$, so that in particular
this allows for spinning M2-branes contrary to the case of single odd-spheres considered here where spinning M2-branes are not possible. 

As also remarked in \cite{Armas:2012bk}, another important next step would be to consider the case in which we have many thermal
 (spinning) giant gravitons moving along the $S^1$ of $S^n$ and taking the limit in which they are smeared along this circle. 
 This would reveal the the difference between the smeared and non-smeared phases at finite temperature, and elucidate the connections with for example
 the superstar \cite{Myers:2001aq}, bubbling $\ads$ solutions \cite{Lin:2004nb} and bubbling $\ads$ black holes \cite{Liu:2007xj}.
A related outstanding question is to examine the connection between our null-wave giant gravitons
 (which have $SO(m-1) \times U(1)$ isometry with $m=5$ for D3 and $m=4$ for M5) and the lower
supersymmetric bubbling  geometries that have been considered in the literature (see e.g. Refs.~\cite{Gava:2006pu}).
In this connection, considering thermal versions of giant gravitons with less supersymmetry \cite{Mikhailov:2000ya} is  expected to be relevant as well. 

We finally point out that the null-wave giant gravitons do no have a counterpart in the usual weakly coupled world volume theory description. It would
be interesting to reconsider this by studying the thermal DBI (or M-brane) theory and exploring an appropriate limit.  This would be worthwhile in view of
finding a precise dual description of the null-wave giant gravitons. More generally, via the AdS/CFT correspondence our thermal spinning giant graviton solutions 
are expected to correspond to a thermal state in the dual gauge theory. It would be very interesting to find a description of this thermal state in the gauge theory and compare its properties to those of the thermal giant graviton, in particular the free energies found in Eq.~\eqref{freeenergy} in the low temperature limit
and the accompanying low/high spin results.

\section*{Acknowledgments}

We thank Matthias Blau, Jan de Boer, Jakob Gath, Troels Harmark and Jelle Hartong
for useful discussions. JA thanks NBI for hospitality. 
The work of JA was partly funded by the Innovations- und Kooperationsprojekt C-13 of the Schweizerische Universit\"{a}tskonferenz (SUK/CRUS).
The work of NO is supported in part by the Danish National Research Foundation project ``Black holes and their role in quantum gravity''.


\begin{appendix}

 
\section{Details on solution space} \label{app:reparameterization}
In this appendix we give further details on the solution space presented in Secs.~\ref{sec:setup}-\ref{sec:thermal} and establish the relation between the results presented in this paper and those obtained in \cite{Armas:2012bk}.

\subsection{Alternative parameterization of solution space}
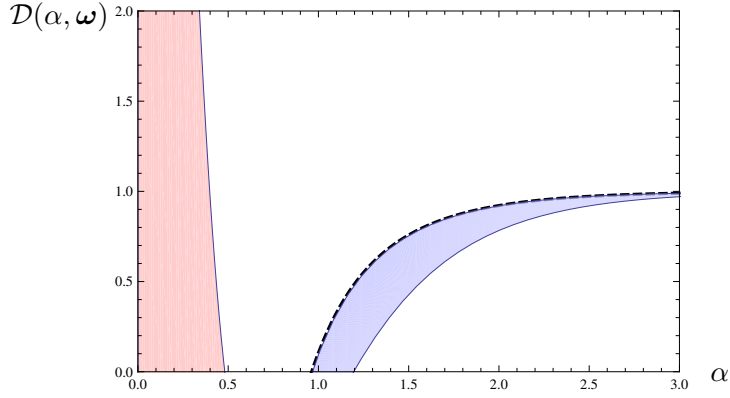
\begin{figure}[!ht]
\centerline{\includegraphics[scale=0.6]{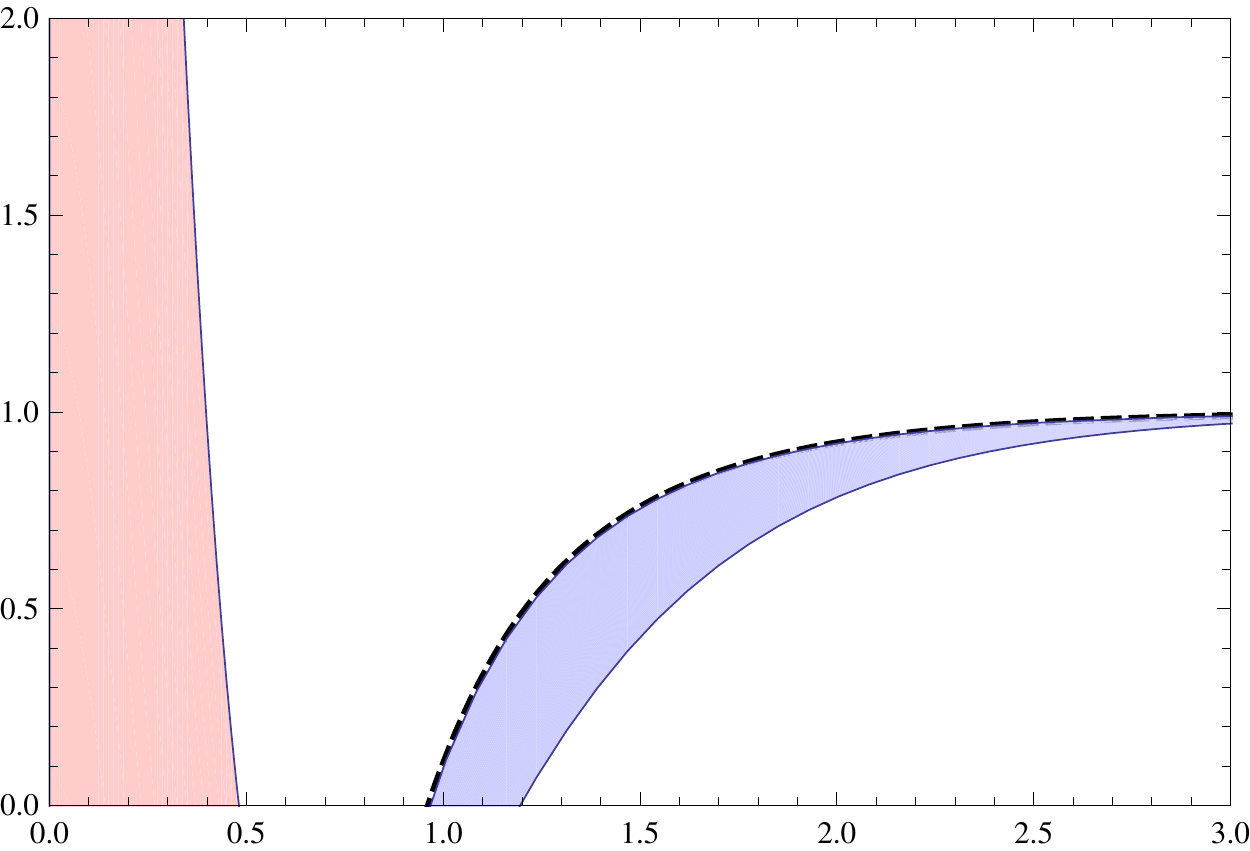}}
\begin{picture}(0,0)(0,0)
\put(335,20){ $\alpha$}
\put(70,155){ $ \mathcal{D}(\alpha,\boldsymbol{\omega})   $}
\end{picture}		
\vskip -.5cm
	\caption{$\mathcal{D}(\alpha,\boldsymbol{\omega})$ as a function of $\alpha$ for $0\le\boldsymbol{\omega}\le1$ and $n=5$~. The dashed black line represents the case $\boldsymbol{\omega}=0$. The vertical axis was restricted to the interval $0\le\mathcal{D}(\alpha,\boldsymbol{\omega})\le2$ while the horizontal axis was restricted to $0\le\alpha\le3$.
	}
	\label{fig:phase}
\end{figure}

Here we reparameterize the equations of motion and solution space of Sec.~\ref{sec:setup} such that the connection with the solution space of the non-spinning thermal giant graviton found in \cite{Armas:2012bk} is more apparent. To this aim, we define a new parameter $\boldsymbol{\omega}$ such that
\beq
\boldsymbol{\omega}=\frac{\omega^2r^2}{\textbf{k}^2}~~.
\eeq
Using this newly defined parameter, the equation of motion \eqref{exteq} can be rewritten as
\beq \label{altext}
\left(n-2+\mathcal{R}_1\boldsymbol{\omega}\right)|k_{\text{w.v.}}|^2+\Omega^2r^2\left(1-\mathcal{R}_1(\boldsymbol{\omega}+1)\right)+(n-1)\Omega r |k_{\text{w.v.}}|\mathcal{R}_2=0~~.
\eeq
For clarity of presentation we focus on the case $n=5$. In this situation Eq.~\eqref{altext} admits the following family of solutions
\beq\label{altsol}
\Omega_{\pm}=\frac{|3+\boldsymbol{\omega}\mathcal{R}_1|}{\sqrt{(3+\boldsymbol{\omega}\mathcal{R}_1)^2L^2-8(1+\Delta_{\pm}(\alpha,\boldsymbol{\omega}))r^2}}~~,
\eeq
where we have defined
\beq
\Delta_{\pm}(\alpha,\boldsymbol{\omega})=-\frac{1}{8}\left(3\mathcal{R}_1+8\mathcal{R}_2^2\pm4\mathcal{R}_2\sqrt{\mathcal{D}(\alpha,\boldsymbol{\omega})}+\boldsymbol{\omega}\mathcal{R}_1(\mathcal{R}_1-4)\right)+\frac{1}{2}~~,
\eeq
with 
\beq
\mathcal{D}(\alpha,\boldsymbol{\omega})=-3(1-\mathcal{R}_1)+4\mathcal{R}_2^2+\boldsymbol{\omega}\mathcal{R}_1(2+\mathcal{R}_1(\boldsymbol{\omega}+1))~~.
\eeq
Indeed, setting $\boldsymbol{\omega}=0$ in Eq.~\eqref{altsol} yields the form of $\Omega_{\pm}$ obtained in \cite{Armas:2012bk} for thermal giant gravitons expanded into the $S^{5}$ part of $\ads_{5}\times S^{5}$. A necessary condition for the solution \eqref{altsol} to exist is $\mathcal{D}(\alpha,\boldsymbol{\omega})\ge0$. In Fig.~\ref{fig:phase} we exhibit the dependence of $\mathcal{D}(\alpha,\boldsymbol{\omega})\ge0$ on $\alpha$ within the range $0\le\boldsymbol{\omega}\le1$.
\vskip .7cm
From Fig.~\ref{fig:phase} we see that there are two regions of possible spinning giant graviton configurations. The black dashed line depicts the case $\boldsymbol{\omega}=0$ obtained in \cite{Armas:2012bk} for which there is only one region of possible solutions. As the spin is increased the solution space is composed of a blue region (Region 1) and of a red region (Region 2). It is possible to determine analytically the ranges of $\alpha$ defining both regions by solving $\mathcal{D}(\alpha,\boldsymbol{\omega})=0$~. This leads to the ranges
\beq \label{ranges}
\begin{split}
\text{Region 1:}&~~~\left(\frac{9}{4}+\boldsymbol{\omega}\right)\le\cosh^2\alpha<\infty~~,~~\boldsymbol{\omega}\ge0~~\\
\text{Region 2:}&~~~1\le\cosh^2\alpha\le\left(\frac{1}{4}+\boldsymbol{\omega}\right)~~,~~\boldsymbol{\omega}>\frac{3}{4}~~.
\end{split}
\eeq
From \eqref{ranges} we see that Region 1 exists for all values of $\boldsymbol{\omega}$ while Region 2 only appears after the spin parameter $\boldsymbol{\omega}$ is increased beyond the value $\boldsymbol{\omega}=3/4$. At the lower bound of Region 1 and at the upper bound of Region 2 the two branches of solutions $\Omega_\pm$ meet each other. Note that Region 2 can be decomposed into a thermodynamically stable and unstable part. The unstable part lies within the range $1\le\cosh^2\alpha\le 3/2$ as it has negative heat capacity \cite{Grignani:2010xm}. For generic $(m,n)$ we obtain similar bounds as in \eqref{ranges}, in particular for the non-spinning case, these are $ 5/3 \le\cosh^2\alpha<\infty$ for the M5-giant graviton and $10/3 \le\cosh^2\alpha<\infty$ for the M2-giant graviton.


\subsection{Range of $\textbf{k}$} \label{app:detailsrange}
The ranges \eqref{ranges} together with charge conservation \eqref{branecurrent} allow to determine the bounds on $\textbf{k}$ mentioned in Sec.~\ref{sec:solution}.
Focusing on $n=5$ and on the lower bound of Region 1 we obtain the bound for $\textbf{k}$
\beq
\text{Region 1:}~~~\hat{T}\thinspace \frac{\left(9+4\boldsymbol{\omega}\right)^{\frac{3}{8}}}{2^{\frac{1}{4}}(3\sqrt{3})^{\frac{1}{4}}\left(5+4\boldsymbol{\omega}\right)^{\frac{1}{8}}}\le\textbf{k}\le 1~~.
\eeq
In the case $\boldsymbol{\omega}=0$ this agrees with the result found for non-spinning giant gravitons in \cite{Armas:2012bk}. For Region 2, the upper bound in \eqref{ranges} allows us to write the bound on the thermodynamically stable part as
 \beq
 \text{Region 2 stable:}~~~\hat{T}\le\textbf{k}\le\hat{T}\thinspace\frac{\left(1+4\boldsymbol{\omega}\right)^{\frac{3}{8}}}{2^{\frac{1}{4}}(3\sqrt{3})^{\frac{1}{4}}\left(4\boldsymbol{\omega}-3\right)^{\frac{1}{8}}}~~,
 \eeq
while for the unstable part it is instead allowed in the entire interval
\beq
 \text{Region 2 unstable:}~~~\hat{T}\le\textbf{k}\le1~~.
 \eeq
For the bounds on $\textbf{k}$ for the stable part of both regions we observe that there is a gap in the allowed values of $\textbf{k}$ for which there does not exist a giant graviton configuration. This is the gap observed in Sec.~\ref{sec:thermal} for the maximal size giant graviton. The same features are observed for the other values of  $(m,n)$.


\subsection{Maximum temperature} \label{app:detailsmaxtemp}
The solution space does not admit configurations at any temperature $T$. As already seen for the non-spinning giant graviton in \cite{Armas:2012bk} there exists a maximum temperature beyond which giant graviton configurations cease to exist. This bound is obtained from the charge conservation equation \eqref{branecurrent} which can be recast into the form
\beq \label{Qcon1}
\textbf{k}^{m-1}=\frac{Q_{(n-2)}G}{\Omega_{(m)}}\frac{4(4\pi)^{m}\mathcal{R}_1(\alpha)\text{cosh}^{m-1}\alpha}{(m-1)^{m}\mathcal{R}_2(\alpha)} T^{m-1}~,
\eeq
where the ratios $\mathcal{R}_1$ and $\mathcal{R}_2$ are  defined in \eqref{R12}. The maximum temperature that the giant graviton can attain in the thermodynamically stable region is obtained from \eqref{Qcon1} when $\cosh\tilde\alpha$ takes the value that gives rise to the lower bound of Region 1 in \eqref{ranges}. Generically, we can define the maximum temperature as
\beq
T_{\text{max}}^{m-1}=\left[\frac{Q_{(n-2)}G}{\Omega_{(m)}}\frac{4(4\pi)^{m}\mathcal{R}_1(\alpha)\text{cosh}^{m-1}\alpha}{(m-1)^{m}\mathcal{R}_2(\alpha)} \right]^{-1}\!\!\!\!\!\!|_{\alpha=\tilde\alpha}~~.
\eeq
For the case of the spinning giant graviton on $\ads_{5}\times S^{5}$ this results in
\beq
T_{\text{max}}=T_{\text{static}}\left(\frac{6\sqrt{3}\sqrt{5+4\boldsymbol{\omega}}}{\left(9+4\boldsymbol{\omega}\right)^{\frac{3}{2}}}\right)^{\frac{1}{4}}~~.
\eeq
From the above expression we see that as the spin parameter $\boldsymbol{\omega}$ increases, the maximum temperature that the giant graviton can attain decreases. This is again a generic feature for any $(m,n)$.


\subsection{The special case $\Omega=\omega$}\label{Omegaequalomega}
Here we analyze the case for which $\Omega=\omega$. This is a peculiar case as it corresponds to a branch of solutions for which there is no continuous limit that connects it with the thermal non-spinning giant graviton of \cite{Armas:2012bk} but it still admits a limit in which it connects to the usual $\frac{1}{2}$-BPS giant graviton. In this situation the spin orbit interaction term in \eqref{exteq} vanishes and the equation of motion can be written as
\beq
(n-2)\left(1-\Omega^2(L^2-r^2)\right)+\Omega^2r^2+(n-1)\Omega r \sqrt{1-\Omega^2(L^2-r^2)} \mathcal R_2=0~~.
\eeq
For clarity of presentation we focus on the case $n=5$ but we note that the above equation admits a solution for any $n$. For $n=5$ the solution takes the form
\beq \label{ow}
\Omega_{\pm}=\frac{3}{\sqrt{9L^2-8(1+\Delta_{\pm}(\alpha))\thinspace r^2}}~~,
\eeq
where
\beq
\Delta_{\pm}(\alpha)=-\frac{1}{2}\left(2\mathcal{R}_2^2(\alpha)\pm\mathcal{R}_2(\alpha)\sqrt{\mathcal{D}(\alpha)}\right)+\frac{1}{2}~~,~~\mathcal{D}(\alpha)=4\mathcal{R}_2^2(\alpha)-3~~.
\eeq
We see that \eqref{ow} allows for two branches of solutions. However, one must remember that the condition $\textbf{k}^2=1-\Omega_{\pm}^2L^2\ge0$ must be imposed, implying $\Omega_\pm\le L^{-1}$. A straightforward check tells us that the upper branch solution always violates this requirement (except in the strict limit $\alpha\to\infty$). Hence we conclude that for the case $\Omega=\omega$ only the lower branch of solutions exists. Imposing the same requirement on the fluid velocity $\textbf{k}$ for the lower branch leads to the allowed range for $\alpha$ in solution space
\beq \label{rangeow}
\frac{9}{8}\le\cosh^2\alpha<\infty~~.
\eeq
This range implies that there is a thermodynamically stable region and an unstable region which ranges from $ 9/8 \le\cosh^2\alpha\le 3/2$. This furthermore means that this branch of solutions does not admit a neutral limit (as one cannot approach $\alpha=1$), i.e., they must be always charged and supported by the background gauge field. Moreover, the range \eqref{rangeow} implies that in both stable and unstable regions, the fluid velocity must satisfy the bound $\hat{T}\le\textbf{k}\le1$. Another interesting feature of this branch of solutions is that both ends of the interval \eqref{rangeow} correspond to zero-temperature limits. The limit $\alpha\to\infty$ corresponds to either the usual extremal limit of Sec.~\ref{sec:setup} or the null-wave limit of Sec~\ref{sec:nullwave}. The limit $\alpha\to 9/8$, using the fact that $\Delta_{-}(9/8)=-1$, implies $\Omega_{-}=L^{-1}$ and hence that $\textbf{k}\to0$. Therefore, by charge conservation \eqref{Qcon1} we see that for the charge $Q_{(n-2)}$ to remain constant we must have $T\to0$. This is another type of null-wave giant graviton configuration but not a regular one since in this limit the thickness $r_0$ remains finite and hence all thermodynamic quantities presented in Sec.~\ref{sec:setup} diverge except for the product $TS$ which remains finite. Further, in this limit the configuration satisfies the relation $\textbf{F}=\textbf{E}-\hat T \textbf{S}=\textbf{J}+\boldsymbol{\mathcal{S}}$~, which is the BPS relation found in Sec.~\ref{sec:nullwave}.


\end{appendix}

\addcontentsline{toc}{section}{References}
\footnotesize

\providecommand{\href}[2]{#2}\begingroup\raggedright\endgroup


\end{document}